 \documentclass[a4paper,8pt]{article}
\pdfoutput=1
\usepackage{jcappub} 
\usepackage[T1]{fontenc} 
\usepackage{graphics}   
\usepackage{hyperref}      
\usepackage{color}  
\usepackage{caption}
\usepackage{subcaption}    
\usepackage{ifthen}                 
\usepackage{amssymb}
\usepackage{mathrsfs}
\usepackage{bbold}
\usepackage{float}                           
\usepackage{multirow}
\usepackage{mathtools}%
\usepackage{cellspace,tabularx}%

\def\dl{{\rm d}\ell}

\def\W{\mathcal W}

\def\be{\begin{equation}}
\def\ee{\end{equation}}
\def\barr{\begin{array}{lr}}
\def\earr{\end{array}}
\def\bea{\begin{eqnarray}}
\def\eea{\end{eqnarray}}

\def\a{\alpha}
\def\b{\beta_s}

\def\s{\sigma}

\def\D{\Delta}

\def\has{{\texttt{Haslam}}}
\def\SV{{\texttt{Stockert-Villa}} }
\def\wmap{{\texttt{WMAP}}}
\def\planck{{\texttt{Planck}}}
\def\BP{{\texttt{BeyondPlanck}}}
\def\mcmc{{\texttt{MCMC}}}
\def\mem{{\texttt{MEM}}}

\def\mg{\textcolor{black}}

\begin{document}
\title{Statistical properties of Galactic synchrotron temperature and polarization maps --- a multi-frequency comparison}

\author[a,b,1]{Fazlu Rahman}{\note{corresponding author}}
\author[a,c]{, Pravabati Chingangbam},
\author[d]{and Tuhin Ghosh}
\affiliation[a]{Indian Institute of Astrophysics, Koramangala II Block,       
  Bangalore  560 034, India}
\affiliation[b]{Department of Physics, Pondicherry University, R.V. Nagar, Kalapet, 605 014, Puducherry, India}
\affiliation[c]{School of Physics, Korea Institute for Advanced Study, 85
Hoegiro, Dongdaemun-gu, Seoul, 02455, Korea}
\affiliation[d]{National Institute of Science Education and Research, An OCC of Homi Bhabha National Institute, Bhubaneswar 752050, Odisha, India}
\emailAdd{fazlu.rahman@iiap.res.in}
\emailAdd{prava@iiap.res.in}
\emailAdd{tghosh@niser.ac.in}
\abstract{Understanding the statistical properties of synchrotron emission from our Galaxy is valuable from the perspective of observations targeting signals of cosmological origin, as well as for understanding physical processes in our Galaxy. 
In this work, we extend the analysis of Rahman et al.~\cite{Rahman:2021azv} to --- (a) all-sky observed maps of total foreground emissions at different frequencies provided by \wmap, \planck~and \texttt{Stockert-Villa}, 
(b) component separated synchrotron temperature maps provided by \wmap, \planck~and \BP, and (c) component separated polarization maps provided by \wmap~and \planck. The tools we use are Minkowski functionals and tensors. 
Our main goals are twofold. First, we determine the variation of morphological properties of the total foreground maps with observing frequency and compare with simulations. This study elucidates how the morphology varies with frequency due to the relative dominance of different foreground components at different frequencies. Secondly, we determine the nature of non-Gaussianity and statistical isotropy of synchrotron fluctuations towards smaller scales using various component separated synchrotron temperature and polarization maps. We find that all maps exhibit kurtosis-type non-Gaussianity, in agreement with the \has~map. This result can be an important input for the modelling of small-scale synchrotron fluctuations for component separation pipelines. This also suggests that residual synchrotron contamination in CMB will manifest as kurtosis and will not be captured by three-point statistics. 
From a comparison of the different component separated maps, we find that \BP~and \wmap~\mcmc-\texttt{e} agree well with \has~at all scales.  The other maps  show differences of varying statistical significance.
Our analysis suggests a combination of residual AME and/or free-free emissions and point sources as contributing to these differences, and underscores the need for further improvement of the pipelines.
} 

\maketitle
\section{Introduction}
\label{sec:intro}

Diffuse synchrotron emission is the dominant component of the total Galactic emissions at radio frequencies and remains a significant component up to microwave frequencies of roughly 100 GHz. In observations that target the measurement of feeble signals of cosmological origin, in particular, the cosmic microwave background (CMB) and 21 cm signal from the Epoch of Reionization (EoR), modelling and removing  synchrotron and other foreground components (free-free emission, thermal dust emission, anomalous microwave emission (AME) etc.) are essential steps. For this reason, it is crucial to understand the spectral and statistical properties of each foreground component and use that knowledge as an input while developing component separation techniques~\cite{Eriksen:2004ss,Delabrouille:2008qd}. 
Small-scale fluctuations of these emissions are often assumed to be statistically isotropic Gaussian random fields (see e.g., \cite{Tegmark:1999ke,Jelic:2008jg,Marthi:2017}), and it is important to test the validity of that assumption with the observed data, in particular, in sky regions at high Galactic latitudes. 

The synchrotron frequency spectrum is known to \mg{follow} a  power-law form $I_{\nu}\propto\nu^{\beta_s}$, $\nu$ being the electromagnetic frequency. The spectral index, $\beta_s$, depends on the energy distribution of cosmic ray electrons,  $N(E)\propto E^{-p}$, as $\beta_s=-(p+3)/2$~\cite{Rybicki1986}. Since the electron distribution shows spatial variations, $\beta_s$ is expected to vary across the sky. Moreover, $\beta_s$ is a function of frequency, with the spectra steepening at higher frequency bands due to the ageing of synchrotron sources, synchrotron self-absorption, presence of multiple synchrotron components etc.~\cite{kogut2012}. 
Given the non-linear interactions and fluctuations of the synchrotron emission sources, the spatial fluctuations of the synchrotron intensity are  expected to be highly non-Gaussian and anisotropic. Non-Gaussianity of synchrotron fluctuations have been studied using higher order statistical tools such as bispectrum ~\cite{Jung2018,Coulton2019}, skewness-kurtosis \cite{Ben-David:2015b,2019MNRAS.487.5814V} and Minkowski functionals~\cite{Rana:2018oft}. These studies have all reported highly non-Gaussian fluctuations at large angular scales and a decrease in the level of non-Gaussianity as we go down to smaller angular scales. In tune with this, it is a common practice in modelling synchrotron emission  to assume that small-scale fluctuations are Gaussian and statistically isotropic~\cite{Waelkens:2008gp,Thorne2017}. It is then very important to quantify the scale at which this assumption becomes valid.

Due to the non-Gaussian nature, the power spectrum, which has been used to study the synchrotron maps~\cite{LaPorta2008,Planck2018IV,Martire2022}, cannot fully capture the statistical properties of these emissions. Therefore, there is a strong need to employ methods that capture information beyond the two-point function. The standard non-Gaussianity estimators such as bispectrum, skewness and kurtosis and other $n$-point functions, when used alone, also fail to fully capture these properties due to lack of a priori knowledge of the true nature of these foreground fields. 
Morphological descriptors such as Minkowski functionals and tensors are well suited to study the statistical features of synchrotron radiation and other foreground components since they are non-local, contain information of all orders of $n$-point functions and are not biased towards any particular type of non-Gaussianity (see e.g. Chingangbam \& Park \cite{Chingangbam:2013}).

This paper is the second in a series where the authors study in detail the morphology of different components of the Galactic emissions in the radio to microwave frequencies. 
In \mg{Rahman et al.}~\cite{Rahman:2021azv}, we carried out a detailed examination of the small-scale non-Gaussianity of the all-sky 408 MHz \has~synchrotron map using Minkowski functionals and tensors. We found that within the limited resolution of the \has~map, the level of synchrotron non-Gaussianity is not low enough to assume Gaussianity, although there is a decreasing trend towards smaller scales. On the other hand, statistical isotropy was found to be valid at \has~resolution. We also found that non-Gaussianity at small scales is of kurtosis-type. 
In this paper, we extend this analysis and focus attention on the morphological properties of  observed synchrotron temperature and polarization maps provided by \wmap~\cite{bennett_nine-year_2013}, \planck~\cite{Planck2018I} and \texttt{BeyondPlanck}~\cite{BP:Main}. For comparison, we also include the combined map of the 1.4 GHz radio continuum survey from Stockert and Villa-Elisa telescopes~\cite{1986A&AS...63..205R,testori_radio_2001}.  
We analyze various all-sky maps observed at frequencies from 1.4 GHz to 33 GHz and make a comparison with \has~map. We also compare with simulated foreground maps given by the Python Sky Model (PySM)~\cite{Thorne2017}. Then we study the synchrotron maps given by different component separation pipelines of \wmap, \planck, and \texttt{BeyondPlanck}.  The main goal of the current work is to understand how the statistical features of Galactic synchrotron emission vary as a function of frequency and to compare the  non-Gaussian and statistical isotropy nature of these observed synchrotron maps with previous results of the \has~map. A similar analysis focusing on synchrotron polarization was carried out in  a recent work by \mg{Martire et al.}~\cite{Martire:2023ytg}.

The paper is organized as follows. Section~\ref{sec:sec2} contains a discussion of the physics of Galactic synchrotron emission along with a summary of their known statistical properties, 
followed by a description of the data sets used in this work. In section~\ref{sec:sec3}, we describe the methodology and data analysis pipeline we adopt. Section~\ref{sec:sec4} contains the results for the morphology of total foreground maps at different observing frequencies and comparison with simulated total foreground maps.  Section~\ref{sec:sec5} contains the results for the component separated synchrotron temperature and polarization maps. We end with a summary and discussion of our results in section~\ref{sec:sec6}. Appendix~\ref{sec:a1}  describes the estimation of instrumental noise for different experimental setups and maps, and appendix~\ref{sec:a2} describes the signal-to-noise ratio of observed total foreground maps from \wmap~and \planck. \mg{In appendix~\ref{sec:a3}, we discuss the possible contribution of instrumental systematics in our results.} Appendix~\ref{sec:a4} discusses the morphology of two composite fields --- synchrotron plus AME and synchrotron plus free-free. \mg{Lastly, the preparation of the polarization mask used in the polarization analysis is outlined in appendix~\ref{sec:a5}}
\section{Physics of Galactic synchrotron emission and data sets used}
\label{sec:sec2}

In this section, we present brief descriptions of the physics of Galactic synchrotron emission and the data sets used in this work. We also describe Gaussian simulations and simulations of foreground emissions given by PySM.

\subsection{Galactic synchrotron emission}
\label{sec:sec2.1}

The interaction of relativistic electrons in the cosmic rays with the Galactic magnetic field results in synchrotron radiation. The intensity of radiation varies across the sky depending on the spatial variations in the number density of cosmic ray electrons ($n_{e}$) and the Galactic magnetic field ($\text{B}_{\bot}$) distribution. Broadly, we can classify the synchrotron sources into two categories: electrons that are trapped inside the magnetic field of supernova remnants and the diffuse electron species in the cosmic rays spread throughout the Galaxy. It is estimated that about 90\% of the synchrotron emission is of \mg{diffuse origin}~\cite{WMAP:1yearFg} and has a power-law frequency spectrum with a spatially changing spectral index~\cite{Orlando2013}. Moreover, synchrotron emission is highly polarized, with theoretical estimates giving about 70\% polarization efficiency~\cite{Rybicki1986}.

Several studies in the literature focus on large-scale spatial fluctuations of synchrotron emission. The spatial variation of the spectral index, which is important to understand the morphology of synchrotron emission at frequencies relevant for CMB experiments, is  studied in \mg{Miville-Deschênes et al.~\cite{miville2008}, Jew \& Grumitt~\cite{2020MNRAS.495..578J} and Fuskeland et al.~\cite{fuskeland_constraints_2021}}. In \mg{Vidal et al.}~\cite{Vidal2015}, \mg{the} authors have used the \has~map and \wmap~low-frequency channels to identify large-scale filaments and studied the polarization properties of the  filaments. The spatial correlations are  studied  using the power spectrum for both intensity and polarization maps~\cite{Baccigalupi2015, Burigana2006, LaPorta2008, Planck2018IV, Martire2022}. Many of these works have used the frequency dependence of the power spectrum  ($C_{\ell}\propto\nu^{2\beta_s}$) to estimate the synchrotron spectral index and also calculated the correlations between synchrotron and thermal dust emissions~\cite{ LaPorta2008, Martire2022,Planck2018XI,Krachmalnicoff2018}.

Theoretical models have been  constructed to account for the observed features of diffused synchrotron emission for developing component separation pipelines in CMB and other radio experiments. The \texttt{GALPROP} code, which models synchrotron emission for given Galactic magnetic field and cosmic ray propagation scenarios and reproduces the \mg{observed} large-scale features, is extensively used in \wmap~and \planck~analysis~\cite{Strong2011,Orlando2013}. Physical explanations for the observed features in the power spectrum and the large-scale loops and spurs \mg{are provided in} terms of the magneto-hydrodynamic turbulence in the interstellar medium~\cite{Cho:2010kw,Lazarian2012,Mertsch:2013pua,Kendel2018}. Python Sky Model (PySM), used in CMB experiments to simulate emissions in the microwave sky, models synchrotron emission assuming a power-law frequency spectrum and using \has~map and \mg{various} spectral index models~\cite{Thorne2017}.

\subsection{Observed data sets Used}
\label{sec:sec2.2}

There are various publicly available data sets for observations of  diffuse emissions in the frequency range of the order of 100 MHz to 100 GHz. For our study here we choose data sets that have full sky coverage.
The particular data sets that are used in our analysis are given below\footnote{Data sets are taken from\\
\hspace{1cm}\url{https://lambda.gsfc.nasa.gov/} \\
\url{https://pla.esac.esa.int/} \\ 
\url{https://beyondplanck.science/} \\}. 

\subsubsection{All-sky temperature maps at different frequencies}
\label{sec:sec2.2.1}

The observed temperature data at specific frequencies in the range 0.408 to 33 GHz  
 that we analyze (see section~\ref{sec:sec4.1}) are the following.
\begin{enumerate}
\item The \has~408 MHz map~\cite{Haslam:1982} is used as the {\it fiducial} map \mg{which other data sets are compared with}. We use the reprocessed version of the \has~map, given by \mg{Remazeilles et al.}~\cite{Remazeilles:2014mba}. 
\item The combined all-sky map at 1420 MHz from the northern sky survey using Stockert 25 m telescope~\cite{1986A&AS...63..205R} and the southern sky survey with 30 m Villa Elisa telescope~\cite{testori_radio_2001}. We refer to this map as \SV map. 
\item  \wmap~K and Ka bands at 23 and 33 GHz, respectively~\cite{bennett_nine-year_2013}. 
\item  \planck~30 GHz map from PR3 data release~\cite{Planck2018I}. 
\end{enumerate}
We will refer to these as {\it frequency maps}. These maps contain all components of Galactic emissions, with the contribution of each component varying with the observing frequency.

\subsubsection{Component separated temperature and polarization synchrotron maps}
\label{sec:sec2.2.2}

We use the synchrotron maps given by different component separation algorithms in \wmap~and \planck~experiments. The temperature maps are listed as follows:

\begin{enumerate}
    \item {\it \wmap~synchrotron map at K-band derived using Maximum Entropy Method (\texttt{MEM})}: \texttt{MEM} is a pixel-based Bayesian model-fitting technique assuming a spectral model for different Galactic components and using external data sets as priors~\cite{WMAP:1yearFg}. Here, the synchrotron spectrum is considered as a power law with spectral index $\beta_s=-3.0$, and the \has~map is used as the external prior template for synchrotron emission. 
    
    \item {\it \wmap~synchrotron map obtained using \texttt{MCMC} technique}: This is the standard MCMC approach of sampling the posterior distribution of the model parameters given the data sets~\cite{WMAP:9year}. The sky model is constructed based on the spectral features of different astrophysical components. Apart from the 5 \wmap~frequency bands, the \has~map is also included in the data analysis. Different \texttt{MCMC} maps are available based on slightly different assumptions in modelling foreground spectra, which are summarised below.
    \begin{itemize}
        \item \texttt{MCMC-c} --- follows a power-law spectrum for synchrotron, free-free, and thermal dust emissions, with the synchrotron and dust spectral indices as free parameters.
        \item \texttt{MCMC-e} --- includes spinning dust emission, but the synchrotron spectral index is fixed to $-3$.
        \item \texttt{MCMC-f} --- similar to \texttt{MCMC-e}, except that the synchrotron spectral index is allowed to have spatial variation, and the frequency spectrum is assumed to be a pure power-law.
        \item \texttt{MCMC-g} --- similar to \texttt{MCMC-f} with the synchrotron spectral index varying as a function of frequency according to Strong et al.~\cite{Strong2011}, instead of a pure power-law.
     \end{itemize}
    
    \item  {\it \planck~synchrotron map derived using \texttt{Commander} pipeline}: \texttt{Commander} involves standard Bayesian formalism of fitting an explicit parametric sky model to the data and computing the joint posterior distribution. These parameters, corresponding to CMB as well as astrophysical components, are then estimated at each pixel by sampling the posterior via Gibbs sampling~\cite{Eriksen:2008}. In addition to \planck~HFI and LFI data, the nine-year \wmap~sky maps and the \has~map are also used as external data sets to disentangle the degeneracy between different low-frequency Galactic components.
    \item {\it \texttt{BeyondPlanck} synchrotron map}: \texttt{BeyondPlanck} is an end-to-end Bayesian analysis technique that jointly carries out instrument characterization, map-making, and component separation, using a single statistically consistent data model~\cite{BP:Main}. It is implemented as \texttt{Commander3}, an extended version of the \texttt{Commander} pipeline, applied to time-ordered data (TOD). In this analysis, \planck~LFI data sets along with \has~and \planck~dust maps are used as input maps.
\end{enumerate}
For polarization studies, we use the synchrotron maps given by \wmap~\texttt{\mg{MCMC-g}} and \planck~ \texttt{Commander} methods. Although other polarized foreground components like thermal dust are negligible in the low-frequency bands of \wmap~and \planck, polarization maps are dominated by noise, and we use component-separated maps to minimize the bias due to it.  

\subsection{Simulated data}

We use two kinds of simulated data for the analysis. Brief descriptions of the simulations  are given below.

\subsubsection{Gaussian Simulations}
\label{sec:sec2.3.1}
To  quantify the non-Gaussianity and anisotropy of a given map whose properties are a priori unknown we need to compare with suitable Gaussian isotropic simulations. In this work we generate these simulations by calculating the angular power spectrum of the given map, and then using it as input for producing Gaussian isotropic maps. The angular power spectra are obtained using the \texttt{NaMaster}\footnote{\url{https://github.com/LSSTDESC/NaMaster}} package,  which corrects for leakage due to masking, instrumental beam, and pixelation effects~\cite{NaMaster:2019}. Although the given map may contain non-Gaussian and anisotropy features, by using its power spectrum as input to produce Gaussian isotropic maps we ignore the non-Gaussianity and anisotropy. We produce 1000 such maps with the \texttt{synfast} function of \mg{the} \texttt{healpy}\footnote{\url{https://healpy.readthedocs.io/}} package. These simulations accurately replicate the given map at the power spectrum level and serve as Gaussian isotropic equivalents of the map being analyzed.

\subsubsection{PySM simulated temperature maps}
\label{sec:sec2.3.2}
The Python Sky Model (PySM) is a publicly available software package\footnote{\url{https://pysm3.readthedocs.io/}} for simulating foreground emissions in intensity and polarization at microwave and sub-mm frequencies~\cite{Thorne2017}. It provides simulations of the different foreground components. Though our analysis in this paper primarily focuses on synchrotron, other foreground emissions, like free-free and AME emissions, are also relevant when we study the observed data at different frequencies. For this reason, below we discuss the models and template maps of the synchrotron, free and AME incorporated in PySM that are relevant to our analysis here. \mg{PySM provides high-resolution maps of foreground emissions well above the resolution of the template maps used. This is prepared by injecting Gaussian isotropic fluctuations into these maps at small angular scales.} For further details, see Thorne et al.~\cite{Thorne2017}. Note that we discuss simulations for temperature only. We do not make use of polarization simulations as we do not carry out an analysis of the frequency variation of polarization maps.

\noindent{\bf Synchrotron simulations}: synchrotron temperature spectrum is modelled as a power-law scaled to the 408 MHz all-sky \has~map using the following relation,
\begin{equation} 
    I_{\nu} (\hat{n})=I_{0.408} (\hat{n}) \Big(\frac{\nu }{0.408}\Big)^{\beta_s  (\hat{n})} \ ,
    \label{eq:Imodel}
\end{equation}
where $I_{\nu}$ is the amplitude of the map at any frequency $\nu$ (in GHz), $\beta_s$ is the spatially varying synchrotron spectral index and $I_{0.408}$ is the amplitude of \has~map. The above equation gives different models based on how $\beta_s$ is modelled. We consider the three main PySM models of $\beta_{s}$ in our analysis, which are explained below:
\begin{enumerate}
\item {\em Model \texttt{s1}}:  a spatially varying $\beta_s$ is assumed based on the all-sky spectral index map `Model 4' of \mg{Miville-Deschênes et al.}~\cite{miville2008}, prepared using \wmap~and \has~synchrotron maps and a model of Galactic magnetic field. The variation of $\beta_s$ with frequency is not considered here.
\item {\em Model \texttt{s2}}: this model takes into account only the spatial steepening of $\beta_s$ away from the Galactic plane; $\beta_s$ is a function of the Galactic latitude $b$, given by the relation  $\beta_s=\beta_{b=0}+\delta_{\beta}\,\sin|b|$, in accordance with \wmap. Again, $\beta$ does not vary with frequency.
\item {\em Model \texttt{s3}}: this model takes into account the curvature of $\beta_s$ above a certain frequency $\nu_c$, given as $\beta_s=\beta_0+C\,{\rm ln}(\nu/\nu_{c})$. The spatial variation is given by $\beta_0$, which follows the same model as in \texttt{s1}. The curvature parameter $C$ is 0.052 with $\nu_c=23$ GHz~\cite{kogut2012}. 
\end{enumerate}

\begin{figure}[t]
\centering
\vspace{-1cm}
\includegraphics[scale=0.21]{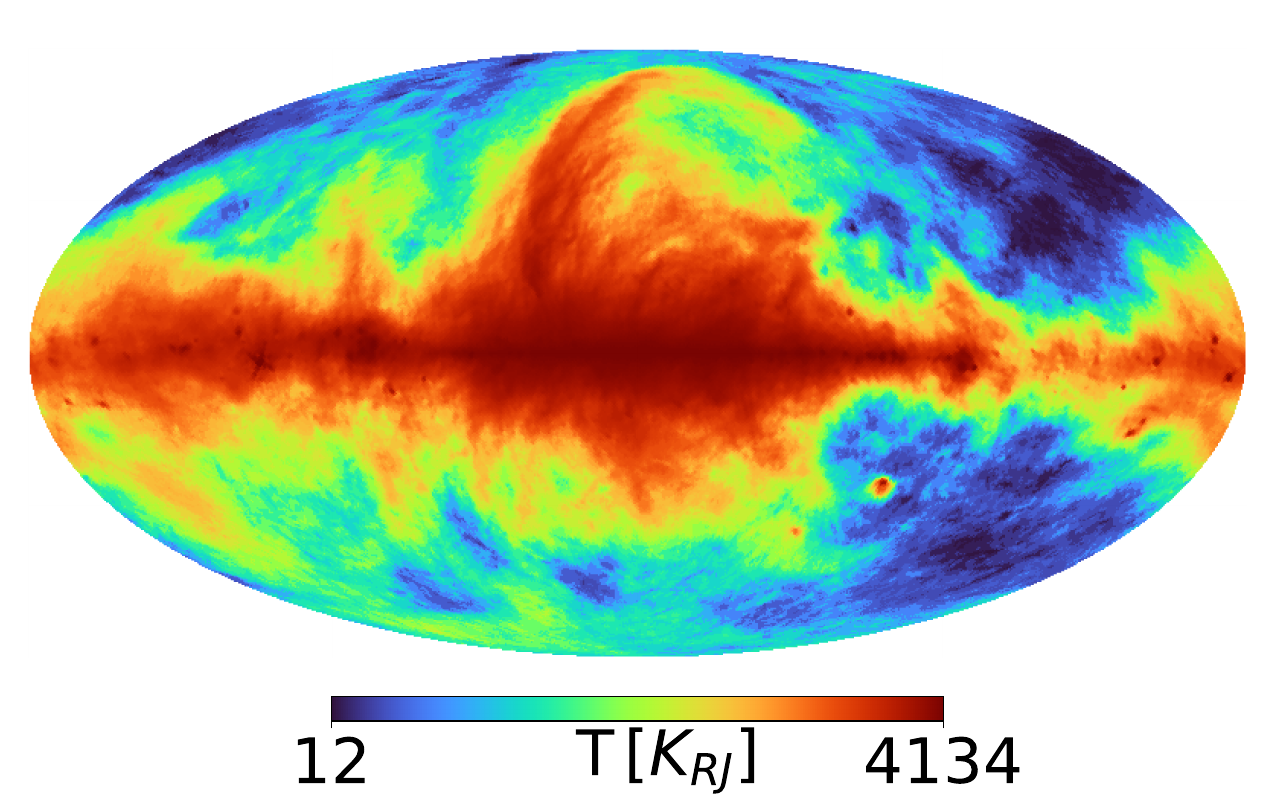} 
\quad \quad
\includegraphics[scale=0.21]{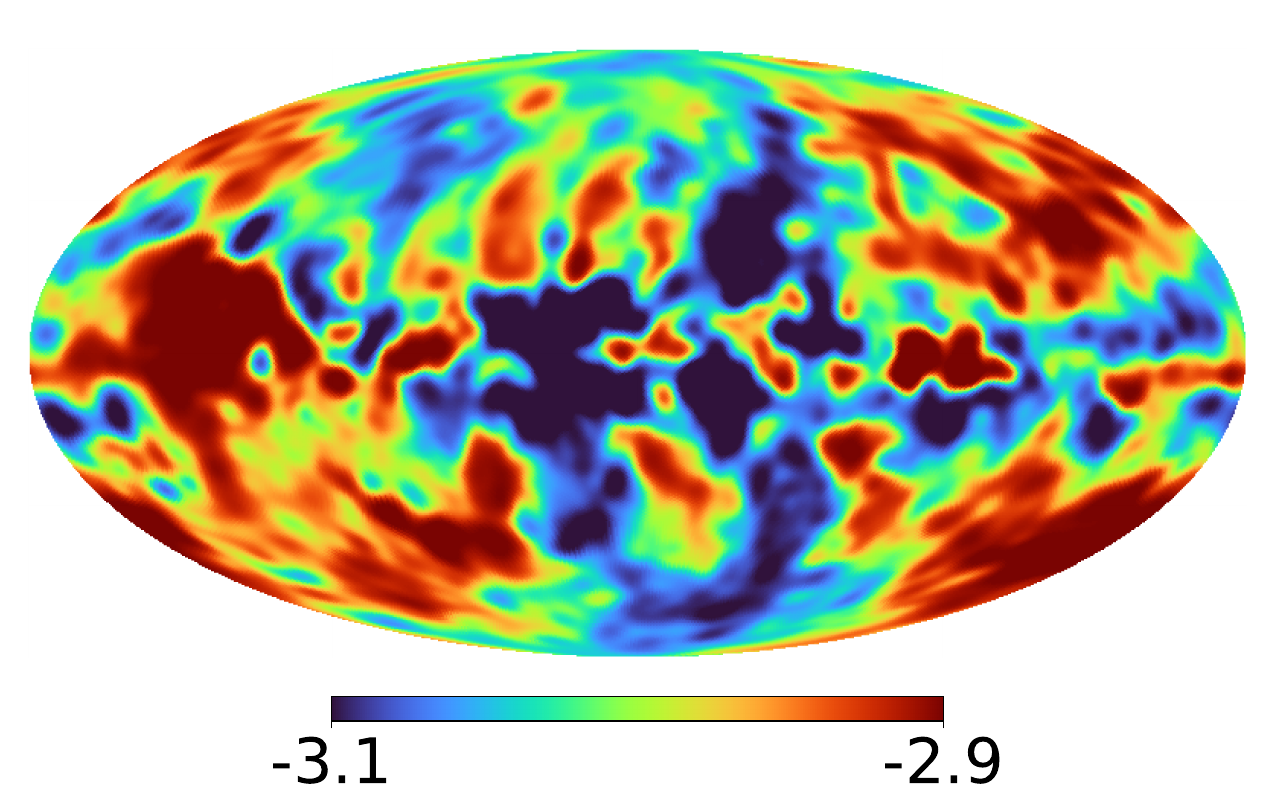}  \quad\quad \includegraphics[scale=0.21]{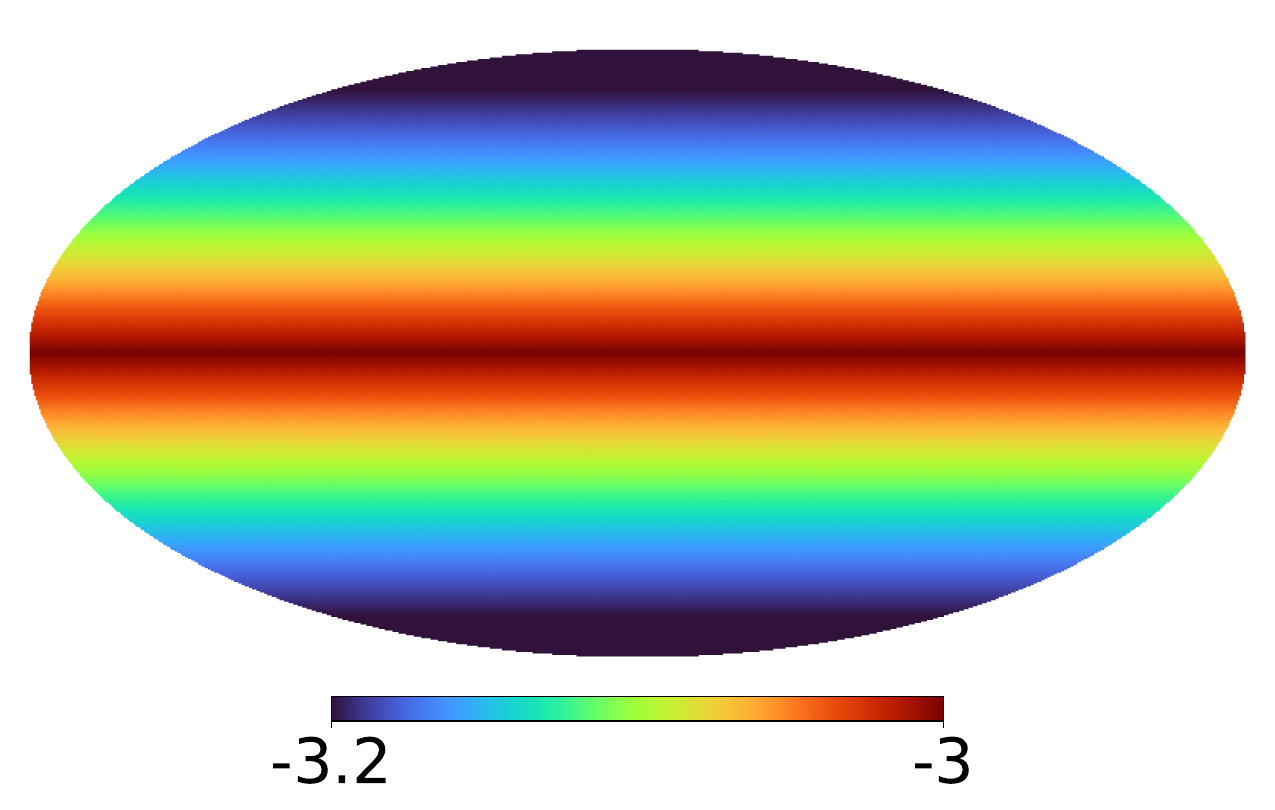}
\caption{{\has~map (left) and} the synchrotron spectral index ($\beta_s$) maps for PySM models \texttt{s1} (middle) and \texttt{s2} (right). The color bars have been histogram equalized.}
\label{fig:pysm_maps}
\end{figure}

Figure
\ref{fig:pysm_maps} shows maps of $\beta_s$ for models \texttt{s1} and \texttt{s2}. Model \texttt{s3} does not differ from model \texttt{s1} in morphology, and hence, we have not shown it. \mg{High-resolution synchrotron maps are obtained by directly adding Gaussian isotropic realizations to the \has~map and then, rescaling it with the spectral index map using eq.~(\ref{eq:Imodel}).}

\noindent{\bf Free-free simulations}: we follow the PySM free-free model called \texttt{f1}. This is the analytical model for free-free emission given in \mg{Draine 2011}~\cite{Drain2011_book} and used in the \planck~2015 \texttt{Commander} analysis~\cite{Planck_2015_X}, which gives a degree-scale map of free-free at 30 GHz. In addition, small-scale fluctuations are added to this map and then rescaled to the desired frequency following a spatially constant spectral index value of $-2.14$.

\noindent{\bf AME simulations}: 
AME model considered is \texttt{a1}, which takes AME as the sum of two spinning dust populations based on the \texttt{Commander} pipeline~\cite{Planck_2015_X}. AME components are characterized by an emission template at a reference frequency and a peak frequency of the emission law. Here, both components follow a spatially varying emission template given by the SpDust2 code. However, the first component has a spatially varying peak frequency, while the second one has a spatially constant peak frequency. Also, small-scale fluctuations are injected into the emission maps. 

\noindent{\bf Dust simulations:} for thermal dust emission, \texttt{d1} model is followed, which assumes the dust spectra to be a single-component modified black body (mbb). \texttt{Planck} 545 GHz map is used as dust templates in intensity, which is then scaled to different frequencies following the mbb spectrum using the (spatially varying) dust temperature and spectral index,  derived from \planck~data using \mg{the} \texttt{Commander} algorithm~\cite{Planck_2015_X}.

\section{Methodology and data analysis pipeline}
\label{sec:sec3}

The statistical tools that are adopted in this work are the Minkowski functionals and tensors, which quantify the morphology of the fields. A brief review of these   morphological methods is presented below, followed by a description of the data analysis pipeline that we follow. 

\subsection{Scalar and tensorial Minkowski functionals}
\label{sec:sec3.1}

This subsection follows \mg{Rahman et al.~\cite{Rahman:2021azv} and Chingangbam et al. \cite{Chingangbam:2017sap,Chingangbam:2021}}. 
Given a smooth random field $u$ on the sphere   $\mathcal{S}^{2}$, the set 
$Q_{\nu_{t}}\equiv\big\{x\in\mathcal{S}^{2}|u(x)\ge\nu_t\s_0\big\}$ 
defines the excursion or level set of the field for the level $u=\nu_t\s_0$, where $\s_{0}$ is the standard deviation of $u$. The iso-field boundaries of $Q_{\nu_t}$, denoted by $\partial Q_{\nu_t}$, form closed contours. 
For each $Q_{\nu_t}$, we focus on the following three rank-two translation invariant tensorial Minkowski functionals,
\bea
{\cal W}_0 (\nu_t)= {\int_{\partial Q_{\nu_t}}} \vec r\otimes \hat n\,{\dl}, \quad 
{\cal W}_1(\nu_t) =\frac14 \int_{\partial Q_{\nu_t}} \hat n\otimes \hat n\,\dl,\quad 
{\cal W}_2(\nu_t) = \frac{1}{2\pi} \int_{\partial Q_{\nu_t}} \hat n\otimes \hat n\, \kappa\, \dl,
\label{MT}
\eea
where 
${\rm d}\ell$ is the line element of $\partial Q_{\nu_t}$, while $\hat n$, $\vec r$, and $\kappa$ are the unit normal vector, position vector and geodesic curvature, respectively,  at each point on $\partial Q_{\nu_t}$. The symbol $\otimes$ is the symmetric tensor product. 
It can be shown that ${\cal W}_0 (\nu_t)\propto \left(\int_{Q_{\nu_t}}{{\rm d}a}\right) \ {\cal I}$, and  ${\cal W}_2 (\nu_t)\propto \left(\int_{\partial Q_{\nu_t}}\kappa\, {{\rm d}\ell}\right) \ {\cal I}$ \cite{mcmullen1997isometry,hug2008space,Schroder2D:2009, Chingangbam:2017sap}, where ${\cal I}$ is the identity matrix and ${\rm d}a$ is the area element of $Q_{\nu_t}$. The factor inside the round bracket for ${\cal W}_0$ is the area integral over the excursion set, which is obtained by converting the line integral in eq. (\ref{MT}) to area integral~\cite{mcmullen1997isometry,hug2008space}.  For ${\cal W}_2$, the factor in round brackets is the curvature integral over the excursion set boundary. 
${\cal W}_1$, however, is not proportional to the identity matrix. Its trace is proportional to the perimeter of the excursion set boundary.

As discussed above, the trace of each of the tensorial quantities in eq.~(\ref{MT}) are proportional to the scalar Minkowski functionals defined below,
\begin{equation} \label{MF}
    V_{0}(\nu_t)=\frac{1}{A}\int_{Q_{\nu_t}}{{\rm d}a}, \quad \quad V_{1}(\nu_t)=\frac{1}{4A}\int_{\partial Q_{\nu_t}} \dl,
    \quad \quad    V_{2}(\nu_t)=\frac{1}{2\pi A}\int_{\partial Q_{\nu_t}}\kappa\, \dl, 
\end{equation}
where $A$ is the area of the space over which the field is defined. $V_{0}$ is proportional to the fraction of area enclosed by the structures in an excursion set, while $V_{1}$ is proportional to the perimeter of the structures per unit area. $V_2$, commonly known as genus, 
is computed as an integral of the geodesic curvature $\kappa$ along the boundary of the structures, per unit area.  For the full sky, $A=4\pi$, while for the masked sky, it is the area of the regions that are not masked. Eqs.~(\ref{MT}) and (\ref{MF}) differ by the area factor. Since $\W_0$ and $\W_2$ are proportional to the identity matrix times the respective scalar MF, they do not provide new statistical information. However, $\W_1$ contains new information  on the directionality and elongation of structures. 

The shape of the MFs and MTs as functions of $\nu_t$ is determined by the nature of the fields, while the amplitudes are determined by their spectral properties.  In particular, analytic expressions for ensemble expectations of MFs and MTs for isotropic Gaussian random fields are well known~\cite{Tomita:1986,Chingangbam:2017sap}. The MFs for a Gaussian isotropic field, having mean zero and normalized by its standard deviation, $\sigma_0$, are given by 
\begin{equation} \label{eqn:GMF}
    V^G_{0}(\nu_t)=\frac{1}{2}{\rm erfc}\left(\frac{\nu_t}{\sqrt{2}} \right), \quad \quad V^G_{1}(\nu_t)=\frac{1}{8\sqrt{2}}\frac{\sigma_1}{\sigma_0}e^{-{\nu_t}^2/2},
    \quad \quad    V^G_{2}(\nu_t)=\frac{1}{4\sqrt{2}\pi^{3/2}}\left(\frac{\sigma_1}{\sigma_0}\right)^2\nu_t e^{-{\nu_t}^2/2}. 
    \end{equation}
These expressions inform us that cosmological information is encoded in the amplitudes related by $\sigma_1/\sigma_0$ of $V_1$ and $V_2$, and the nature of the field is encoded in the functional shapes of the three MFs. 

\noindent{\bf Effect of translation of field values on MFs and MTs}: for any field, translating the field values as $u\rightarrow u-a$, where $a$ is some constant, merely translates the field levels, but does not change the geometrical and topological properties of the field. Subtracting the mean value of the field is such an operation.

\noindent{\bf Effect of scaling of field values on MFs and MTs}: a rescaling $u \rightarrow u/a$, where $a>0$ is some constant, also remaps the field levels. However, this transformation does not alter the topology and the geometry of the excursion sets. We will use this property in sections~\ref{sec:sec4} and \ref{sec:sec5} when we compare the morphology of different fields.

\noindent{\bf Measure of non-Gaussian nature of a given field}: to quantify the non-Gaussian deviation of a given field, we define:
\begin{equation}{\label{eq:DeltaVk}}
    \Delta V_{k}=V_{k}-V_{k}^{G},
\end{equation}
where the superscript $G$ refers to the Gaussian expectation. For a given field whose nature is apriori unknown, we calculate $V_k^G$ from Gaussian simulations obtained from the angular power spectrum of the field (see section~\ref{sec:sec2.3.1}).  The field is said to be mildly non-Gaussian if  $\Delta V_{k}/V_k^G < 1$. Analytic expressions for MFs of mildly non-Gaussian fields are also well known~\cite{Matsubara:2003,Matsubara:2011,Matsubara:2020,Gay:2012}. The expressions are given as perturbative expansions in powers of $\sigma_0$ about the Gaussian expectations given in eq.~(\ref{eqn:GMF}). We do not reproduce the expressions here, but the interested reader can refer to the above references.

\noindent{\bf Test of statistical isotropy}: 
to obtain information about statistical isotropy, we utilize ${\cal W}_1$. Let $\Lambda_{1}$ and $\Lambda_{2}$ be its eigenvalues. Then the ratio, 
\begin{equation}
    \alpha=\frac{\Lambda_{1}}{\Lambda_{2}} \quad \quad  \Lambda_{1}<\Lambda_{1},
   \label{eq:alphadef}
\end{equation}
provides a measure of the relative alignment of iso-field contours and hence quantifies the statistical isotropy of the field. $\alpha \to 1$ implies that the field has statistical isotropy. Deviation from unity indicates that the structures are statistically oriented towards a particular direction.

Our numerical calculation of MFs and MTs follows \mg{Schmalzing et al.~\cite{Schmalzing:1998} and Chingangbam et al.~\cite{Chingangbam:2017sap}}. 
\subsection{Data analysis pipeline}
\label{sec:sec3.2}


For meaningful comparison between different observed data sets, and also meaningful comparison of observed data with simulations, we first need to ensure that the observed and simulated data 
are produced at the same resolution, have the same physical units and all processing steps of masking and bandpass filtering must be identically applied.  The observed data considered in this work are given at different resolutions. We downgrade all the maps (and upgrade \wmap~\texttt{MCMC} maps) to the same resolution given by \texttt{Healpix} resolution parameter, $N_{\rm side}=128$ because the \has~data is given at this resolution.
For consistency, we also convert the maps that are given in units of CMB temperature (${\rm K}_{\rm CMB}$) to the Rayleigh-Jeans unit (${\rm K}_{\rm RJ}$).

The main data analysis pipeline prior to the calculation of MFs and MTs consists of bandpass filtering to remove large-scale modes, followed by applying a mask. These steps are identically applied to all the observed and simulated data. Below we describe the bandpass filtering and masking processes.  

\noindent{\bf Bandpass filtering}: as done in our previous study~\cite{Rahman:2021azv}, to probe  small-scale features of the field, a bandpass filter is applied in the harmonic space to suppress the large-scale fluctuations. The filter function we use is
\begin{equation}
f(\ell)=\frac{1}{4}\left\{1+ {\rm tanh}\left(\frac{\ell-\ell_{c}}{\Delta \ell}\right)\right\}\left\{1-\text{tanh}\left({\frac{\ell-\ell^{*}}{\Delta \ell}}\right)\right\}.
\end{equation}
This filter function contains two cutoff scales.  The lower cutoff, $\ell_c$, determines the scale beyond which larger size (smaller $\ell$) fluctuations are filtered out.  The higher cutoff, $\ell^{*}$,  is set by the beam size of the \has~map (56 arcmins). We use $\ell^*=180$.
We also need to ensure that spurious ringing effects are not introduced after harmonic transform due to a sharp change of $a_{\ell m}$ values at $\ell_c$. A too small value of $\Delta\ell$ will introduce ringing, while too large will not provide a reasonable filter of modes. We choose $\Delta\ell =10$ as a reasonable compromise between the above  two factors. Directly applying the filter on foreground maps results in ringing structures due to the bright Galactic emission. To avoid this, we initially mask the Galactic region by identifying the pixels where \has~map has values greater than 80 K.

\begin{figure}[t]
\centering
 \includegraphics[scale=0.33]{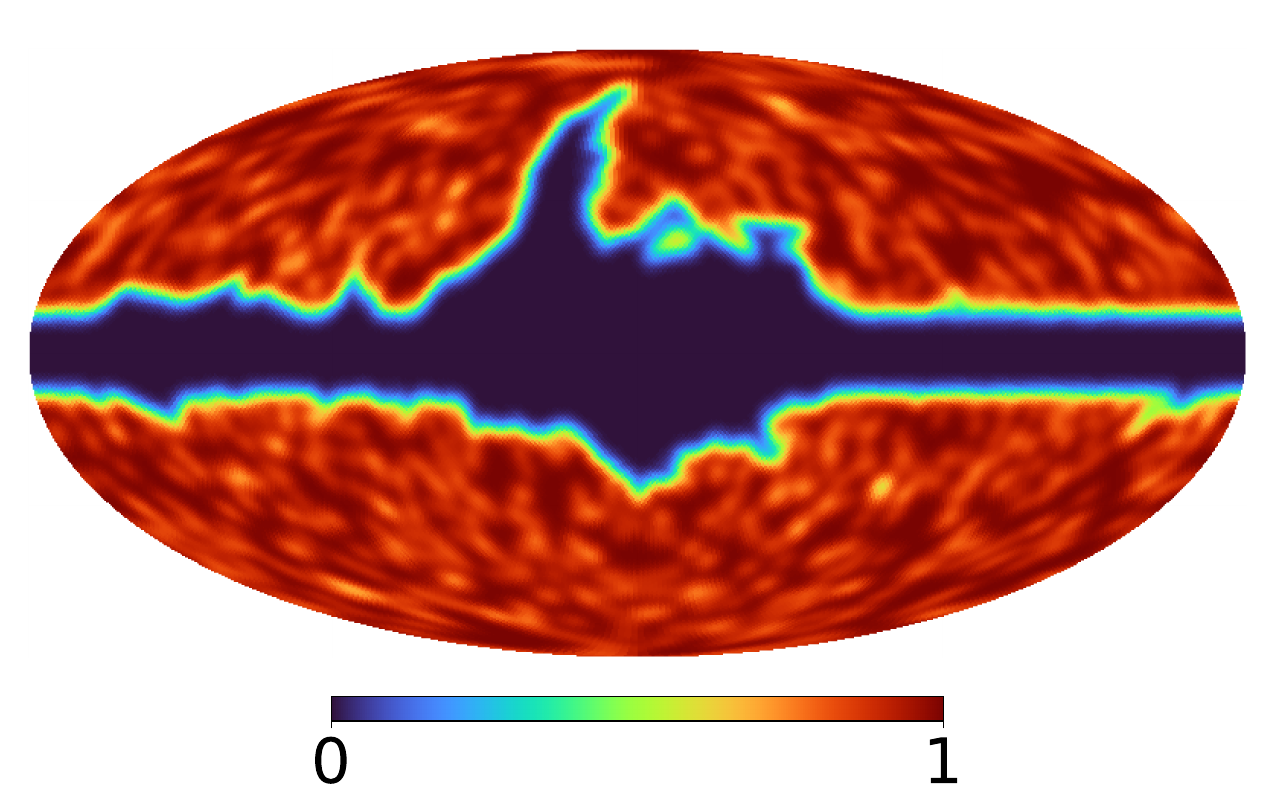}  \quad\quad  \includegraphics[scale=0.33]{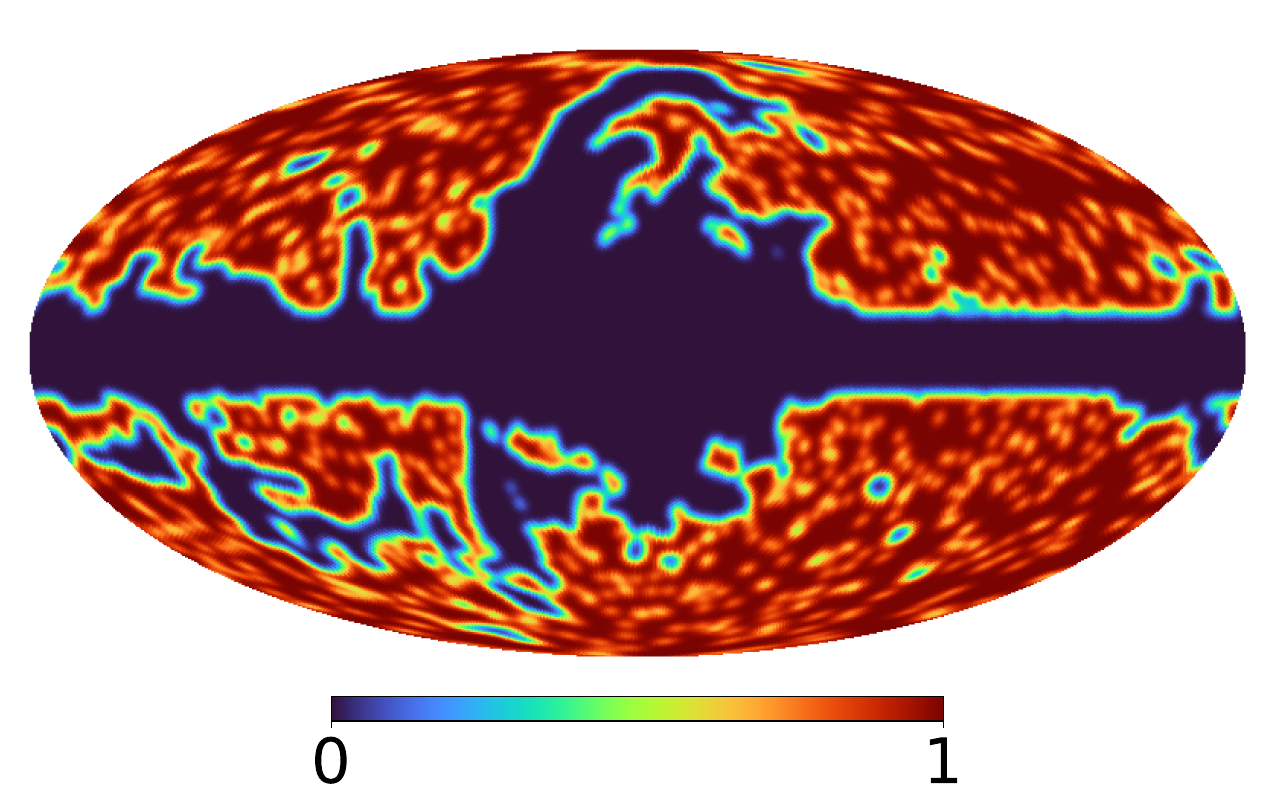}  
 \caption{Maps of the {\it threshold mask} (left) and {\it filament mask} (right), apodized with a Gaussian beam of ${\rm FWHM}=180$ arcmin.} 
\label{fig:mask_bp}
\end{figure}

\noindent{\bf Masking}: we mask some parts of the sky to exclude the brightest regions in the synchrotron sky while retaining the regions where synchrotron signals are significant and where CMB analysis is carried out.

We use the \has~map as the reference for deciding high-emission regions. We also remove point sources identified by \planck. The steps involved in preparing the mask \mg{are as follows}.
\vspace{-.3cm}
\begin{enumerate}
\item Apply a Galactic cut $ \mid b\mid< 10^{\circ}$, $b$ being the Galactic latitude. 
\vspace{-.2cm}
\item Sort all pixels in increasing order of the values of the \has~map, and then mask 25\% of the pixels counting down from the highest value. This corresponds to masking all the pixels in the reference map with values above 35 K. 
\vspace{-.2cm}
\item Lastly, we multiply the above mask by the \planck~ LFI point sources mask to remove contamination from  extra-galactic point sources.
\end{enumerate}
\vspace{-.3cm}
We refer to this final mask as the \textit{threshold mask}. It retains 65\% of the sky. A plot of this mask is shown in the left panel of figure \ref{fig:mask_bp}. 

Another mask that we use for our calculations in order to test the robustness of the results is the so-called \textit{filament mask}~\cite{Vidal2015}. This mask removes large filamentary structures seen in the \has~map and \wmap~polarization maps 
in addition to the masked regions of the threshold mask.  It has a sky-fraction of 53\%. A plot of this mask is shown in the \mg{right} panel of figure \ref{fig:mask_bp}.

To minimise the error due to the sharp mask boundary when performing harmonic transforms, we apodized the masks with a Gaussian smoothing of $3^\circ$ FWHM. Also, we estimate the MFs only on pixels where the apodized mask has values greater than 0.9 so as to avoid any residual contamination from the boundary pixels.

\section{Results --- morphology of observed frequency maps and comparison with PySM simulated maps}
\label{sec:sec4}

In this section, we focus on understanding the angular power spectra and morphology of the observed frequency maps described in  section~\ref{sec:sec2.2.1}. We compare them with PySM simulated maps of the {\em total Galactic foregrounds} consisting of synchrotron, free-free, AME and thermal dust emissions.
We then include a discussion of only synchrotron temperature simulations from PySM.

Out of the observed frequency maps studied here, \has~and \SV maps are known to be predominantly synchrotron emission, while \wmap~and \planck~frequency maps contain CMB, AME, and free-free emissions apart from synchrotron~\cite{Davies:2005za, Ghosh:2012}. From \wmap~and \planck~frequency maps, we subtract the best-fit cleaned CMB maps provided by the respective experiment so as to focus only on the total Galactic foreground emission. We do not subtract the CMB component from the \SV map as the CMB contribution is expected to be negligible.

\subsection{Angular power spectra of observed frequency maps and  PySM simulated total foreground maps}
\label{sec:sec4.1}

\begin{figure}[t]
\begin{center} 
\hspace{-0.5cm}`
\includegraphics[scale=0.53]{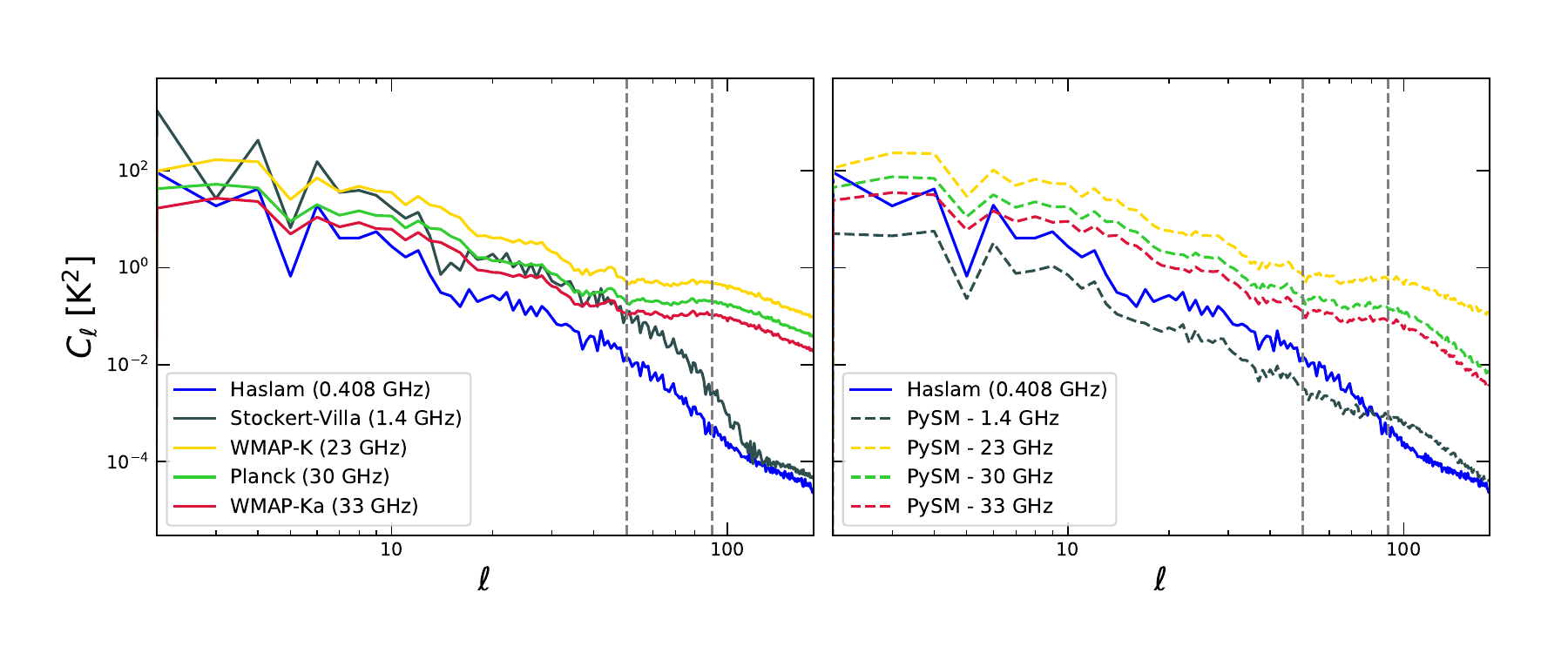}\\
\vspace{-0.5cm}
\includegraphics[scale=0.49]{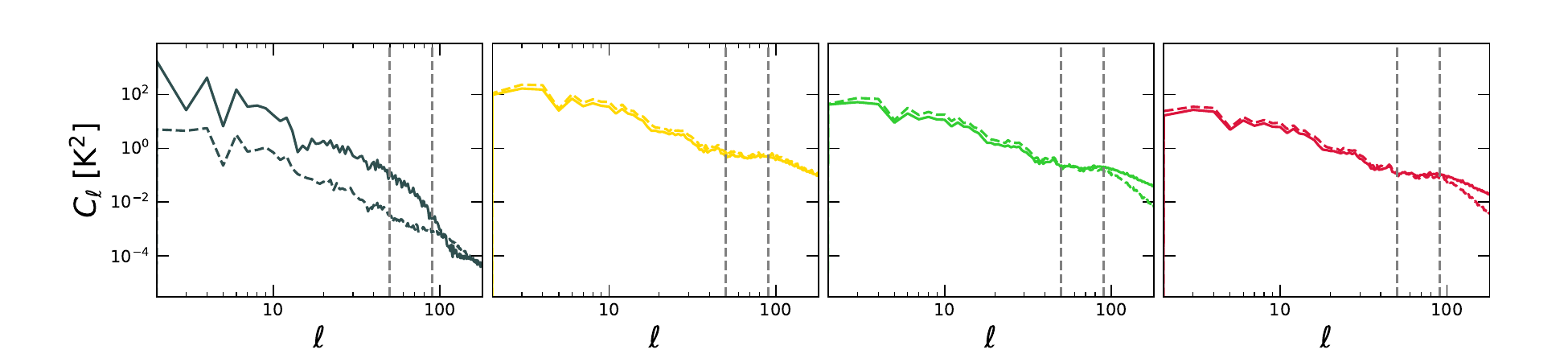}    
\end{center}
\vspace{-0.5cm}
 \caption{{\it Top left:} $C_{\ell}$s of the observed frequency maps. {\it  Top right}:  $C_{\ell}$s of the PySM simulated total foreground maps at the same frequencies as the left panel. {\em Bottom}: Comparison plots of observed frequency (solid lines) and corresponding PySM total foreground maps (dashed lines) at each frequency (increasing frequency from left to right). The vertical lines correspond to $\ell=50,90$.}
\label{fig:Cl_freq}
\end{figure}

For computing the power spectra, all maps are scaled to \has~ frequency (408 MHz) using the power-law synchrotron spectrum given by eq.~(\ref{eq:Imodel}), with a constant spectral index ($\beta_{s}$). For \SV~and the corresponding PySM map, we use $\beta_s= -2.5$, while for other maps at higher frequencies, we use $\beta_s=-3.0$, accounting for the spectral steepening at higher frequencies. We then calculate the angular power spectra of the maps after masking,  using \texttt{Namaster}. This constant scaling makes the amplitudes of the power spectra comparable. But it will not affect the morphology, as mentioned in section~\ref{sec:sec3.1}.  The plots for the observed frequency maps are shown in figure~\ref{fig:Cl_freq} (top left). Also shown are the power spectra of simulations of the total Galactic foreground obtained from PySM at the same frequencies (top right) as the observed frequency maps. From the top left panel, we see that the power spectra at 23, 30 and 33 GHz are distinctly flatter than those of \has~and \SV from roughly $\ell\sim 50$. Apart from a shift of the amplitudes, their shapes are similar. \SV~map shows similar shape of the power spectrum as \has~map till roughly $\ell\sim 90$.  On the top right panels, we observe that the PySM simulated total foreground maps at the three higher frequencies also exhibit similar flattening behaviour as for the observed maps, while 1.4 GHz map is quite different from \texttt{Stockert-Villa}. \mg{We can observe that the shape of power spectrum for 1.4 GHz map is similar to the power spectra of PySM maps at higher frequencies.}

In the bottom panels of figure~\ref{fig:Cl_freq}, we compare $C_{\ell}$ of each observed frequency map with the corresponding simulated map so as to highlight their similarities or differences.  We see that \SV shows considerable differences across all scales from the simulation, as noted above. At 23 GHz,  the power spectra of observed and simulated maps are overlapping, while at 30 and 33 GHz, the observed maps show higher power towards small scales. We will find this comparison of the power spectra between observed and simulated maps useful when we discuss the morphology of the maps below.

\subsection{Morphology of observed frequency maps and comparison with PySM simulated total foreground maps}
\label{sec:sec4.2}

\begin{figure}[t]
\hspace{-.4cm}  \includegraphics[scale=0.53]{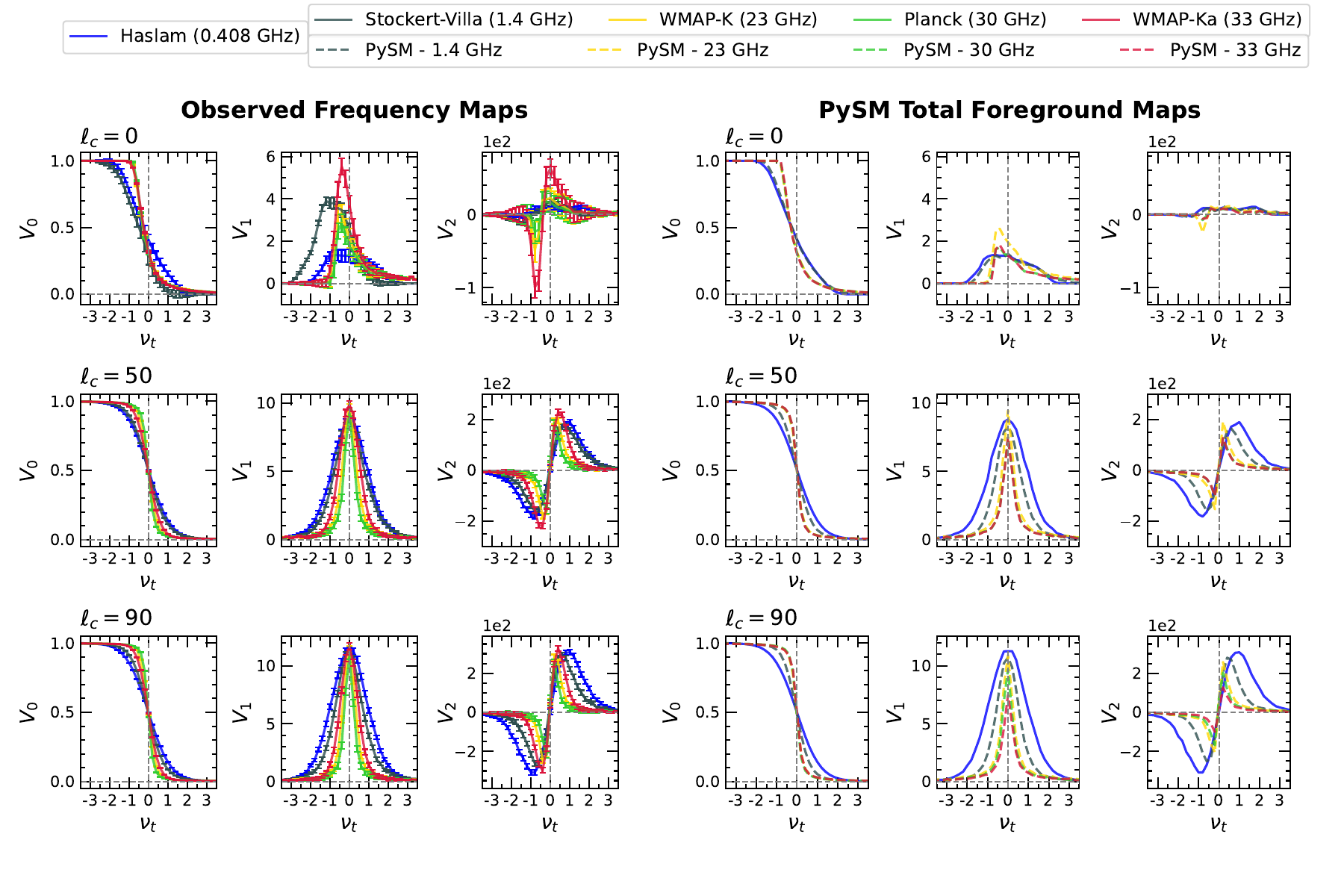}\\
\vspace{-1.2cm}
 \caption{{\it Left, `Observed Frequency Maps'}: MFs versus threshold values, $\nu_{t}$, for each observed frequency map are shown, for different angular scales: $\ell_{c}=0$, 50 and 90. {\it  Right, `PySM Total Foreground Maps'}: Same plots as the left panel, but for PySM total foreground maps simulated at the frequencies corresponding to the observed frequency maps.}
\label{fig:MFs_data_freq_fullcomp}
\end{figure}

MFs are computed for all the maps after bandpass filtering for different values of $\ell_c$ and then masking. Our goal here is to understand how the morphology of the total foreground emissions transforms with frequency --- from the frequency of \has~map where synchrotron dominates to higher frequencies where AME and free-free emissions become the dominant components. This exercise can shed light on how morphological properties of synchrotron emission can be biased by contamination by residual AME or free-free emissions.

To estimate the uncertainties of the MFs arising from cosmic variance and instrumental noise for each observed frequency map 
we compute Gaussian simulations using its angular power spectrum as the input, as described in section~\ref{sec:sec2.3.1}. To each simulation, we add a white noise map specific to the instrument, prepared based on the steps outlined in appendix \ref{sec:a1}. We then compute MFs from the resulting maps and calculate their standard deviation $\s$ at each threshold value. For \planck~map, we use 300 FFP10\footnote{Full Focal Plane simulations~\cite{Planck_2018III_HFI}} simulations, and thus only 300 Gaussian simulations are used. In all plots, we will show $2\sigma$ error bars.

MFs for the frequency maps for $\ell_c=0,\,50$ and 90 are shown on the left panels of figure \ref{fig:MFs_data_freq_fullcomp} under the heading `Observed Frequency Maps'. 
The case $\ell_c=0$ corresponds to no high-pass filtering, meaning that all scales within the range $0<\ell\lesssim180$ are taken into account. 
We find considerable variation with frequency of all three MFs that can be easily discerned by eye, for all $\ell_c$ values. For $\ell_c=0$, there is a systematic increase of the amplitudes of $V_1$ and $V_2$ as the frequency increases. Since $V_2$ is a direct count\footnote{Strictly speaking $V_2$ is equal to the Euler characteristic which is the difference between the numbers of hot spots and cold spots on {\it flat space}. On the sphere,  $V_2$ can be equated to the Euler characteristic to good approximation provided the number of hot/cold spots is high, as is the case for the fields under consideration here.} of the difference between the number of hot spots and cold spots (or structures) this amplitude increase indicates that the number of small scale fluctuations increases as the frequency increases. This increase of power for $\ell>50$ for higher frequencies is observed in the power spectra shown in figure~\ref{fig:Cl_freq}.  
In comparison, \SV map exhibits anomalous behaviour (or non-conformity with the trend followed by \wmap~and \planck~maps) which is most evident from the broad shape of $V_1$. \mg{This could be arising from the poorly understood instrumental effects and calibration uncertainties associated with this map. We intend to address these issues in the future.}

For $\ell_c=50$ and 90, the amplitudes of both $V_1$ and $V_2$ are larger than those of $\ell_c=0$. This is because filtering out large $\ell$ mode leaves behind only smaller-scale structures, which become more numerous as $\ell_c$ increases. We observe that the amplitudes become comparable for all the frequencies,  while the shapes still exhibit considerable differences.  
Visually, we can also see that the shapes  $V_1$ and $V_2$ for higher frequencies are significantly more non-Gaussian than \has~map. This can be discerned by comparing with the expected shapes for Gaussian fields given by eq.~(\ref{eqn:GMF}). This is due to AME and free-free becoming the dominant components at these frequencies. This is demonstrated in  appendix~\ref{sec:a4} where we compare the morphology of AME, synchrotron and AME+synchrotron, and free-free, synchrotron and free-free +synchrotron using simulations obtained from PySM.%

We mention below two factors that can additionally contribute to the  variation of the morphology of the total foreground with frequency and comment on their importance. 

\begin{enumerate}
\item {\it Noise and instrumental systematics}: The error bars shown in figure~\ref{fig:MFs_data_freq_fullcomp} take into account white noise as mentioned earlier. In appendix~\ref{sec:a2} we have determined the contribution of noise to the frequency maps of \wmap~and \planck. The signal-to-noise (SNR) of the maps is found to be much higher than one at all pixels (see figure~\ref{fig:snr_maps}). This implies that noise in the maps has minimal effects on the variation of the morphology with frequency. We have also checked this expectation by repeating all calculations after applying an additional smoothing to all the maps since smoothing has the effect of decreasing noise, though at the cost of losing resolution. We find that our results are robust. Hence, we conclude that the frequency variation of the MFs  is not due to white noise. \mg{Further, the contribution of instrumental systematics has been explored in appendix~\ref{sec:a3}, using \wmap~and \planck~individual year maps. Our analysis shows that the error bars due to instrumental effects are smaller compared to the morphological distinction in these maps. This implies that the systematics play a minimal role in the observed morphological differences.} Note that we have not checked the noise and instrumental effects for the Stockert-Villa map since the data that is available is not sufficient to estimate the SNR of the map and the instrumental uncertainties.  
\item {\it Unresolved extra-Galactic point sources}: While the point sources mask provided by \planck~is included in the threshold mask that we used in our calculations, it is possible that unresolved sources contribute to, and contaminate the observed frequency maps. The effect of this on the angular power spectra will be \mg{to increase the power at} higher multipoles (see also figure 14 and section 4.1 of \mg{Remazeilles et al.}~\cite{Remazeilles:2014mba}). This will lead to an increase in the number of structures at the corresponding scales, which positively contribute to $V_1$ and $V_2$.  
A proper quantification of the contribution, and their distinction from true frequency variation of the total foreground field, will require realistic modelling of the point sources and we postpone it to a future investigation.
\end{enumerate}

Next we focus on simulations of  the total Galactic foreground obtained from PySM at the same frequencies as the observed frequency maps. We then compare (visually) the morphology between the observed and simulated maps. This comparison serves the purpose of checking how well the foreground models of PySM reproduce the observed frequency maps beyond the power spectrum.  We produce simulated maps containing synchrotron, free-free, AME and thermal dust emissions using PySM. The modelling of these components has been described in section~\ref{sec:sec2.3.2}. 

MFs are computed for these maps after identically masking and bandpass filtering as done for the frequency maps. The results are shown in the right panels of figure~\ref{fig:MFs_data_freq_fullcomp}. The values of $\ell_c$ and all the $x$ and $y$ axis ranges are the same as those for the  observed frequency maps for easy comparison. 
For $\ell_c=0$, visually we see that the observed frequency maps and simulated PySM maps at each frequency, except 1.4 GHz corresponding to Stockert-Villa, show good agreement for $V_0$ both in amplitude and shape. The amplitudes of $V_1$ and $V_2$, however, are much lower  for the simulations for all the frequencies. For smaller scales $\ell_c=50$ and 90, we obtain better agreement of the shape and amplitudes of the three MFs with the corresponding MFs of the observed frequency maps. This indicates that PySM reproduces observed foreground maps better towards smaller scales compared to large scales.  

\subsection{Morphology of PySM simulated synchrotron temperature maps}
\label{sec4.2}

\begin{figure}[t]
\hspace{-.4cm}  \includegraphics[scale=0.53]{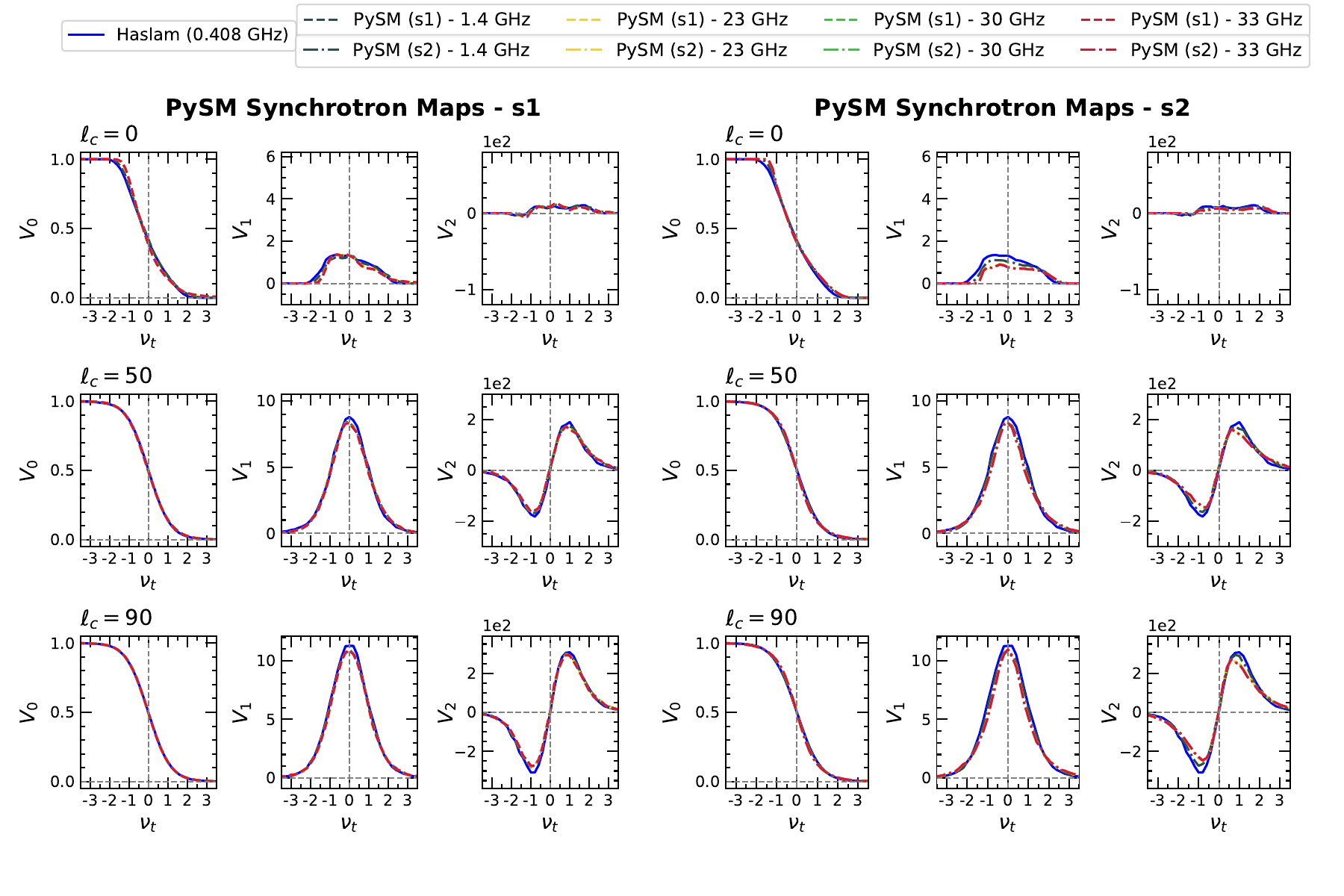}\\    
\vspace{-1.2cm}
 \caption{ Plots of the MFs  versus the threshold $\nu_{t}$ for PySM synchrotron maps for models \texttt{s1} (left panels), and \texttt{s2} (right panels), for $\ell_c=0,50$ and 90. The frequencies of the maps and the scales of all the panels are the same as in figure~\ref{fig:MFs_data_freq_fullcomp}. }
\label{fig:MFs_PySM_synchro_models}
\end{figure}

We now discuss the morphology of PySM simulated synchrotron maps based on the two models \texttt{s1} and \texttt{s2}. We generate the maps at the same frequencies as the observed frequency maps.  The morphology of these maps will be determined by the spatial variation of the synchrotron spectral index.  For model \texttt{s1}, $\beta_s$  fluctuates spatially only mildly. If we denote by $\sigma_0^{\b}$ and $\sigma_1^{\b}$ the rms of $\b$ and the rms of the gradient of $\b$, respectively, we obtain the typical angular fluctuation size of $\b$ to  be $\theta_c=\sigma_0^{\b}/\sigma_1^{\b} \sim 4^{\circ}$. Below this scale, MFs will only encode the morphology of the Gaussian small-scale fluctuations injected in PySM. 
Model \texttt{s2} has even less fluctuations since it varies only along the Galactic latitudes. Therefore, for both models, we expect only a small variation with frequency.

The results  for $V_k$ are shown in figure \ref{fig:MFs_PySM_synchro_models}. All panels follow the same format as figure~\ref{fig:MFs_data_freq_fullcomp}. The  $y$-axis ranges  for $V_k$ are also the same.  As anticipated in the paragraph above, for all three values of $\ell_c$, we find that the variation with frequency of $V_k$ for both the synchrotron models are smaller compared to what we found in figure~\ref{fig:MFs_data_freq_fullcomp}. When we compare these results with the morphology of PySM total foreground maps the large difference can clearly be attributed to the presence of the other foreground components, namely, free-free and AME, in the total foreground.

Further comparing models \texttt{s1} and \texttt{s2}, we find that both models do not differ much. At large scales, model \texttt{s1} varies a bit more from the Haslam morphology than \texttt{s2}. However, towards smaller scales model \texttt{s2} demonstrates slightly larger variation with respect to the \has~map, and the difference increases as the frequency increases. \mg{This may not be immediately apparent when looking at figure~\ref{fig:MFs_PySM_synchro_models}, but it becomes noticeable in the overall trend of average morphology shown in figure \ref{fig:MFs_data_freq_var_fullcomp} of the next subsection.}
\subsection{Comparison of average morphology}

\begin{figure}[t]
\begin{center}
\includegraphics[scale=0.69]{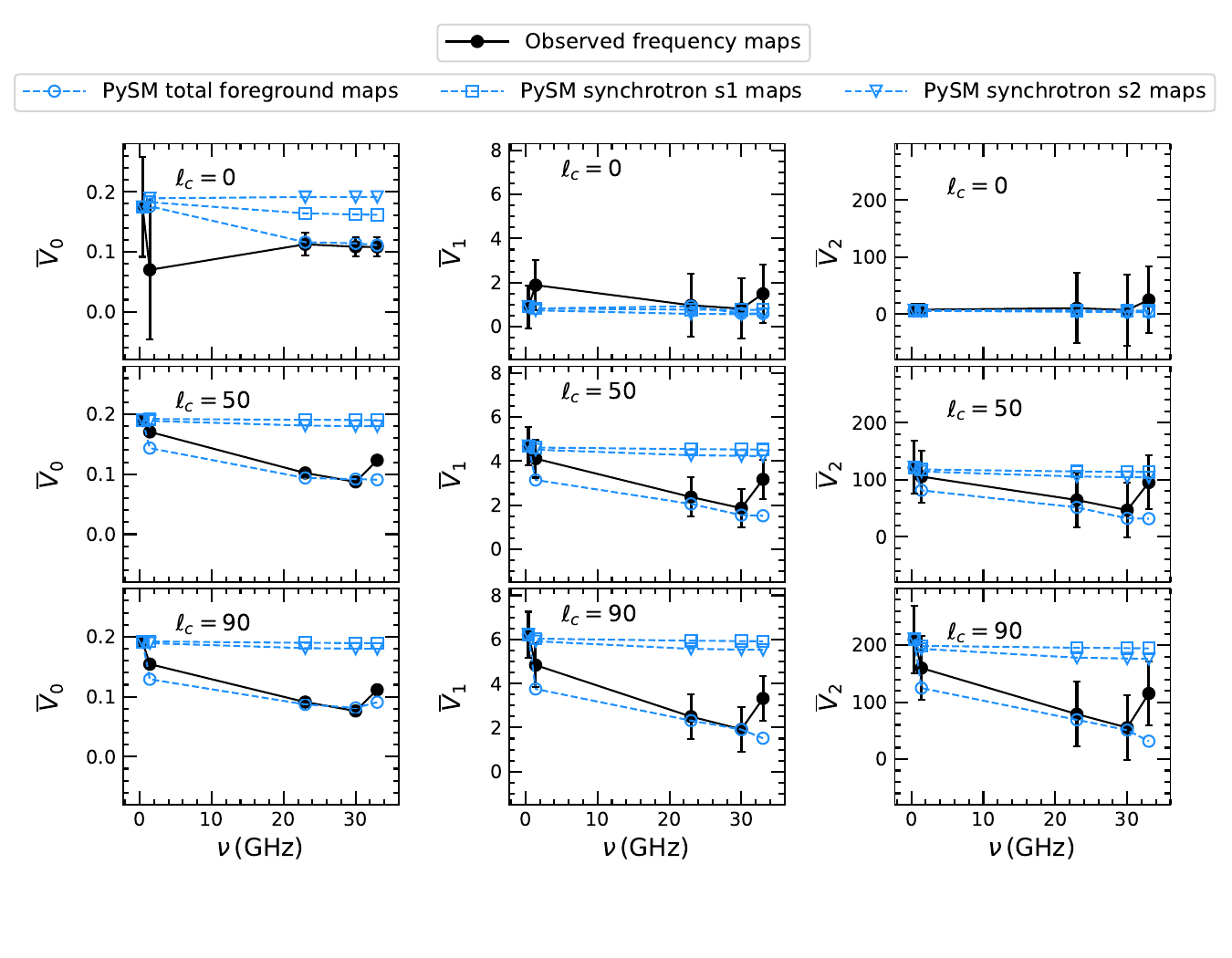}
\end{center}
\vspace{-1.3cm}
 \caption{Average MFs, $\overline{V}_{k}$ defined by eq.~(\ref{eq:V_avg}), for observed frequency maps, PySM simulated total foreground maps, and PySM synchrotron model \texttt{s1} and \texttt{s2},  shown as functions of frequency $\nu$.}
\label{fig:MFs_data_freq_var_fullcomp}
\end{figure}

To visualize how the {\it average} morphology changes as a function of the observing frequency, we define the average value of each MF, $\overline{V}_k$, for each map by integrating over the threshold in the range $\nu_{t,{\rm min}}$  to $\nu_{t,{\rm max}}$, as, 
\be \overline{V}_k (\nu) = \frac{\int_{\nu_{t,{\rm min}}}^{\nu_{t,{\rm max}}} {\rm d}\nu_t V_k(\nu_t,\nu)}{\nu_{t,{\rm max}}-\nu_{t,{\rm min}}}.  
\label{eq:V_avg}
\ee
We choose  $\nu_{t,{\rm min}}=-2$ and $\nu_{t,{\rm max}}=2$ for $V_1$ and $V_2$. For $V_0$,  $\nu_{t,{\rm min}}=0$ and $\nu_{t,{\rm max}}=2$. The average MFs are computed for observed frequency maps and the PySM  maps discussed in the previous two subsections.  The results are shown in figure \ref{fig:MFs_data_freq_var_fullcomp}. The error bars shown for the observed frequency 
maps are $2\sigma$, where $\sigma$ is the sum in quadrature of the standard deviations of the MFs at all the threshold values considered\footnote{We have ignored the correlations of the MFs at different threshold values.}. The values of $\overline V_k$ are higher for increasing $\ell_c$, as expected.     For all three MFs and for all three values of $\ell_c$, we see good agreement between observed frequency maps and  PySM total foreground maps at 23 and 30 GHz. At 33 GHz all plots show large disagreement except $\overline V_0$ at $\ell_c=0$. The PySM synchrotron maps for both models exhibit very tiny variation with frequency.%

Lastly, we stress that the analysis here is focused mainly on synchrotron emission. A full discussion of the morphology of individual foreground components, their contributions to the total morphology and the effect of point sources is beyond the scope of this work and is postponed to future study.

\section{Results --- morphology of component separated synchrotron temperature and polarization maps}
\label{sec:sec5}

Having probed the morphology of observed frequency maps containing synchrotron, free-free, AME and thermal dust emissions at different observing frequencies, we next focus attention on component-separated synchrotron temperature and polarization maps. 

\subsection{Synchrotron temperature maps}
\label{sec:sec5.1}

\noindent

Different component separation pipelines used in various CMB experiments utilize distinct methodologies as outlined in section~\ref{sec:sec2.2.2}. 
In an ideal situation where all pipelines give the same true synchrotron map,  all the maps should exhibit the same morphology. In reality, the \mg{effectiveness} of separating out the different astrophysical components differs, and there can be residual contamination from other components in the component maps. Comparison of the different component separated maps is usually made at the map level by examining the scatter between pixel values or at the power spectrum level (see e.g., Planck Collaboration XXV~\cite{Planck2015XXV}). Here we compare using MFs, thereby, going beyond the power spectra to compare the integrated effect of all orders of n-point correlations. 

\subsubsection{Power Spectra of synchrotron temperature maps }
\begin{figure}[t]
\begin{center}
\includegraphics[height=9cm,width=16.3cm]{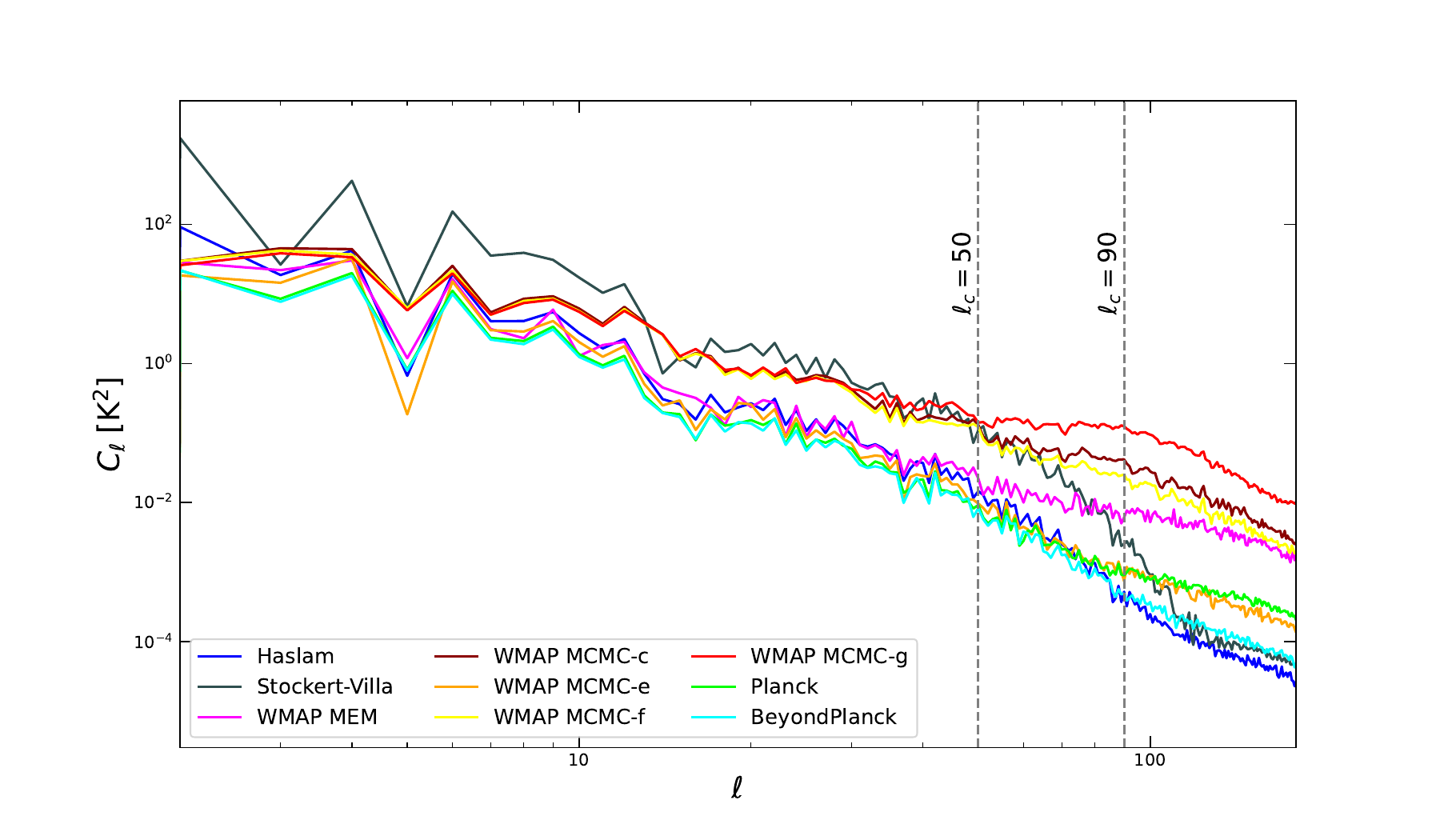}\\
\vspace{-0.48cm}
 \hspace{-.4cm}
 \includegraphics[height=5cm,width=16.3cm]{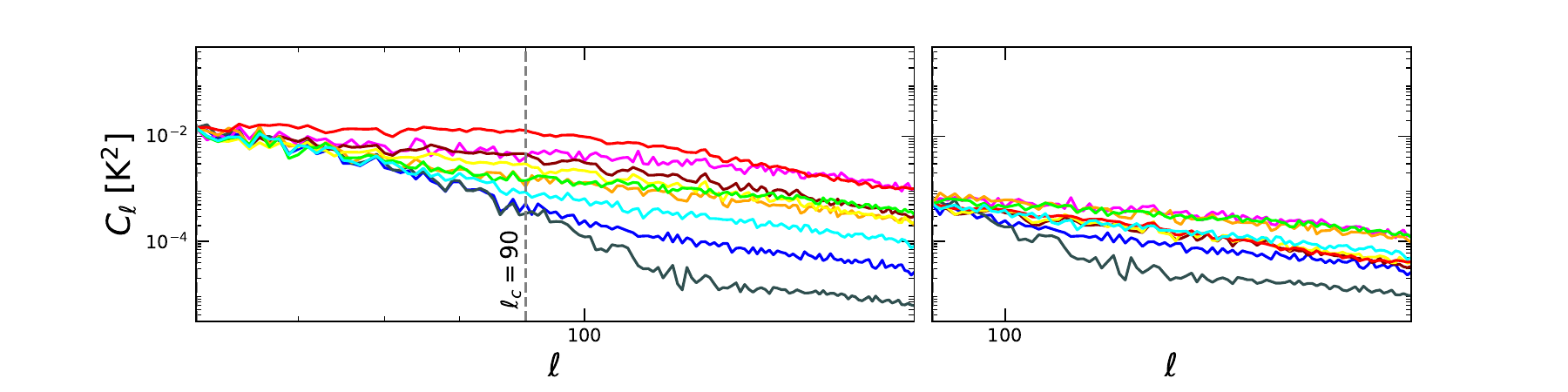}
\end{center}
\vspace{-0.7cm}
 \caption{{\it Top}: Angular power spectrum, $C_{\ell}$, for the full multipole range $0\le \ell \le 180$, of different component separated synchrotron temperature maps given by \wmap~and \planck. $C_{\ell}$ for \has~and \SV maps are also given for comparison. {\it Bottom left}: Same plot as the top but for multipole range $\ell=50$ to 180. The amplitudes of all maps have been rescaled to match that of \has~map at $\ell=50$ so as to highlight the slope differences. {\it Bottom right}: Same plot for multipole range $90\le \ell\le 180$, and amplitudes rescaled to match that of \has~ map at $\ell=90$.  }
\label{fig:power_spectrum}
\end{figure}
We begin our analysis by discussing the angular power spectra ($C_{\ell}$) of the various component separated synchrotron maps estimated using \texttt{NaMaster}. $C_{\ell}$s for \has~and \SV maps are also included for comparison. Among the different synchrotron maps, \wmap~maps are provided at a reference frequency of 23 GHz, while \planck~and \BP~maps are given at 408 MHz. For a meaningful comparison of different maps at the power spectrum level, we have re-scaled all the maps to the Haslam frequency, using a constant spectral index value, $\beta_s=-3.0$ for \wmap~and $\beta_s=-2.5$ for \SV maps. As discussed in section \ref{sec:sec3.1}, this frequency scaling does not alter the overall morphology of the maps under study.
 
The power spectra are shown in figure \ref{fig:power_spectrum}.  
The top panel shows the full multipole range of interest, $0\le \ell\le 180$. We observe that at low $\ell$ (large angular scales), the power spectra of all the synchrotron maps roughly follow a power-law form, with different spectral indices.  \BP~map shows the closest agreement with \has~map across all scales. \wmap~\texttt{MCMC-e} and  \planck~also track the behavior of \has~map down to relatively smaller scales ($\ell\sim 80$) compared to the other \wmap~maps.   
 From this behaviour, we can anticipate that these three maps will show morphological properties that are relatively closer to \has~map.  Note that $C_{\ell}$ of \wmap~\texttt{MCMC-g} (and possibly \texttt{f}) exhibit discernible similarity with  $C_{\ell}$s of \wmap~K and Ka with a flattening of power at $\ell\gtrsim 50$ (compare with figure~\ref{fig:Cl_freq}).  
The power spectrum of  \SV~map is quite different from the others. It exhibits a `knee' with a change of slope and  becomes relatively flat above $\ell\sim 100$.

The bottom left panel of figure \ref{fig:power_spectrum} shows the multipole ranges  $50\le \ell\le 180$ (left), where we have rescaled the amplitudes for all maps to match that of \has~map at $\ell=50$. This rescaling is done so as to highlight the differences in the shape of the power spectra. The $y$-axis range is the same as the top panel. Towards larger $\ell$ (small scales), relative to \has~map, we observe varying degrees of flattening of the power spectra of the different maps (except \texttt{Stockert-Villa}). As mentioned in section~\ref{sec:sec4.2} there can be some contribution of residual point source contamination to this flattening of power. 
Next, the bottom right panel of the same figure shows the multipole ranges  $90\le \ell\le 180$ (left), with the amplitudes for all maps rescaled to match that of \has~map at $\ell=90$. We still observe some flattening, though of a lesser degree, of the power spectra of the different maps relative to \has~map.

\subsection{Minkowski functionals  for synchrotron temperature maps}

\begin{figure}[t]
\hspace{-1.2cm}\includegraphics[scale=0.58]{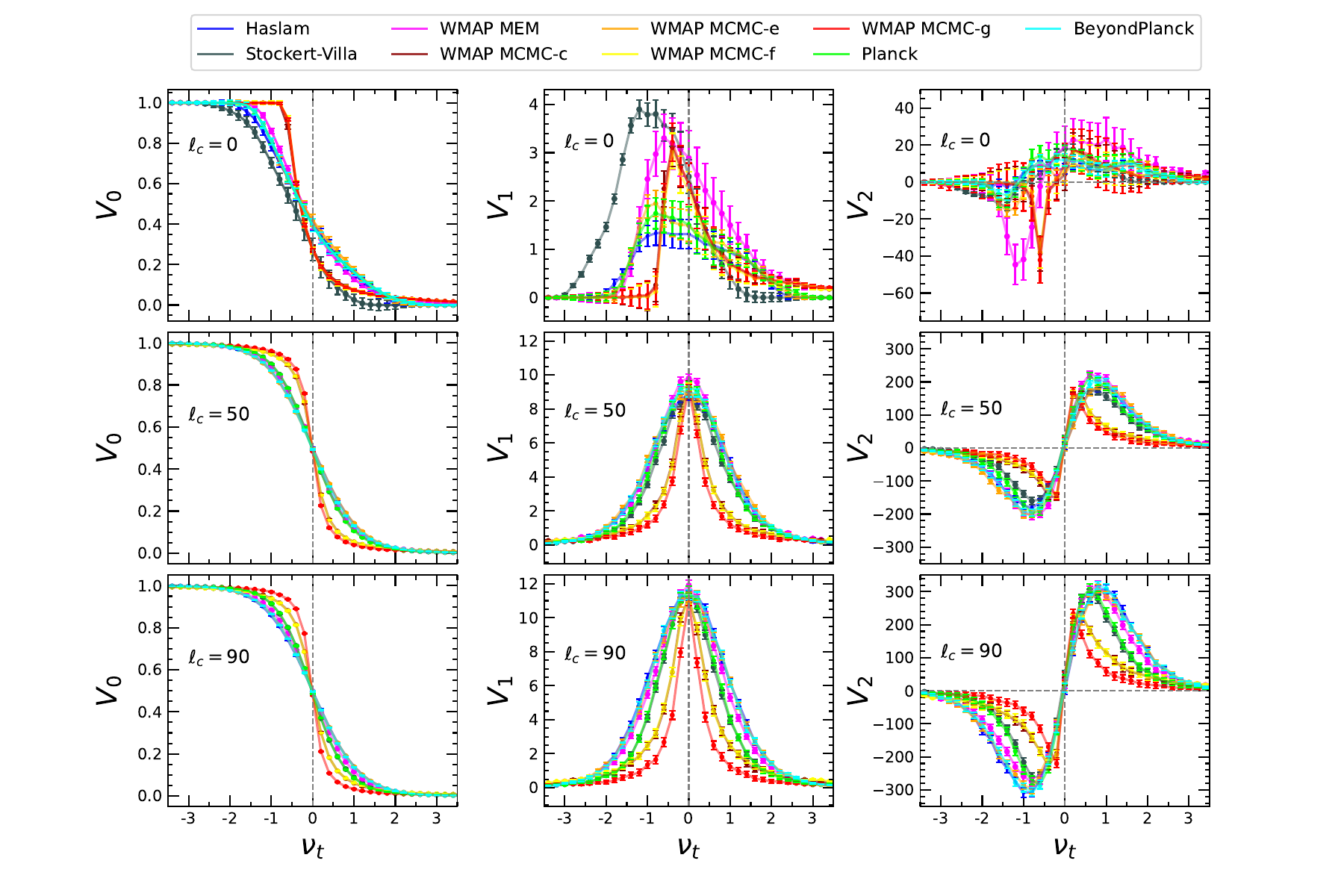}  
\vspace{-0.9cm}
 \caption{Scalar MFs for various component separated synchrotron maps for $\ell_{c}=0$ (top row), $\ell_{c}=50$ (middle row), and $\ell_{c}=90$ (bottom row).}
\label{fig:MFs_data}
\end{figure}

Figure \ref{fig:MFs_data} shows the MFs for all the component separated maps under consideration.  The error bars shown are $2\sigma$, computed from Gaussian simulations to which white noise maps are added (see appendix \ref{sec:a1}). 
The top row corresponds to $\ell_{c}=0$ and hence, is a comparison of the morphology of the full maps containing information on the  entire range of scales.  It is evident that \texttt{BeyondPlanck}, \planck, and \wmap~ \texttt{MCMC-e} maps exhibit good agreement with \mg{the} \has~map for all MFs. This correlates with the agreement of their power spectra at most scales.  Since \has~map is used as a template for  \planck~and \texttt{BeyondPlanck} maps the agreement is not surprising. The other maps are significantly different from \has~map at different threshold ranges, for all three MFs. \wmap~\texttt{MCMC-c, f, g} show good agreement with each other, while \wmap~\texttt{MEM} show significant difference from them. \SV is also significantly different.

Next we examine the MFs for intermediate to small scales set by $\ell_c=50$, shown in the second row of figure \ref{fig:MFs_data}. From all three MFs, we find that  \texttt{BeyondPlanck} and \wmap~\texttt{MCMC-e} agree with \has~map within 2$\sigma$. The MFs of these maps are also closest to Gaussian behaviour, as can be inferred from a comparison of their shapes with eq.~(\ref{eqn:GMF}). From $V_1$ we see that \planck~map differs from \has~at roughly $2\sigma$, with a marginal  increase of non-Gaussianity, while the other MFs show good agreement.  \SV also shows similar behaviour as \planck~for $V_0$ and $V_2$.  
\wmap~\texttt{MCMC-c, f, g} again show significant differences from \has~map and exhibits much higher level of non-Gaussianity. They show good agreement amongst themselves though with \texttt{g} differing from \texttt{c,f} at $2\sigma$.  The similarity of the shapes of the MFs for these three maps with those of the composite fields AME+synchrotron, and free-free+synchrotron, shown in appendix~\ref{sec:a4} is noteworthy. Combined with the similarity of the power spectra of these fields with those of the total foreground mentioned in the previous subsection, this suggests that there is contamination by residual AME and/or free-free emissions in these maps.
Interestingly, the filtering of large scales modes $\ell< 50$ renders  \wmap~\texttt{MEM} map closer to \has~map.

The MFs for small scales set by $\ell_c=90$ are shown in the last row of figure~\ref{fig:MFs_data}. From all MFs, we find that  \texttt{BeyondPlanck} and \wmap~\texttt{MCMC-e} still show good agreement with \has~map. The difference of \planck~and \SV is now much more significant compared to when large-scale modes are included. \wmap~\texttt{MCMC-c, f, g} again show significant difference from \has~map with high level of non-Gaussianity.  The difference of \texttt{g} from  \texttt{c,f} is much more pronounced. As discussed above for $\ell_c=50$, this indicates the presence of residual AME and/or free-free emissions in these maps. \wmap~\texttt{MEM} still shows good agreement with \mg{the} \has~map at roughly 2$\sigma$.

To summarize, we find significant morphological differences between \has, \BP, \planck~and \wmap~\mg{maps}. Our analysis suggests that there are two possible physical causes for these differences. We discuss them below. 
\begin{enumerate}
\item {\em Unresolved point sources}: It has been shown in \mg{Andersen et al.}~\cite{BP:Int_Fg}  that \BP~is contaminated by fewer unresolved point sources compared to \planck.  (See figure 22 of \mg{Andersen et al.}~\cite{BP:Int_Fg} which displays difference maps between the \planck~and \texttt{BeyondPlanck} foreground component maps where the presence of unresolved point sources in \planck~maps at high latitudes are clearly visible).  From this, combined with the flattening of the power spectrum of \planck~map towards small scales relative to \has~map, we deduce that the presence of point sources is responsible (at least partly) for the difference of the MFs between \planck~and \has~at small scales\footnote{We have also computed the MFs for the synchrotron temperature map provided by the \texttt{Cosmoglobe} joint {\it WMAP-LFI} analysis that was recently made public~\cite{Watts:2023vdc}. Our analysis shows that with respect to morphology, \texttt{Cosmoglobe} synchrotron maps are consistent with those of \texttt{BeyondPlanck} maps.}. Since we use the same point sources mask for the \wmap~maps also, there must also be some contribution of residual point sources to their morphology. \mg{The similarity of \texttt{MCMC-e} to \has~ is interesting because it is the one that assumes a spatially constant power-law form for the intensity in the synchrotron modelling. Other \texttt{MCMC} methods keep the spectral index as a free parameter. Given that the point sources also follow a power-law spectra  (albeit with a steeper index), \texttt{MCMC} maps, except \texttt{e}, are more susceptible to erroneously identifying point sources as synchrotron signals.}
\item {\em Contamination by residual AME and/or free-free emissions}: 
A major improvement in the \planck~component separation methods over \wmap~is the better estimation of AME through the improved model for AME spectra \cite{Planck2015XXV}. So the morphological similarity amongst the \texttt{MCMC-c, f,} and \texttt{g} maps and the difference from \has, \planck, and \texttt{BeyondPlanck} indicate that residual AME component contributes to the overall morphology of these \texttt{MCMC} maps. However, our analysis does not distinguish contamination by residual AME from free-free; there can also be contributions from free-free (see the shapes of the MFs in appendix~\ref{sec:a4}). 
\end{enumerate}

\subsubsection{Nature of non-Gaussianity of component separated synchrotron maps}
\label{sec:sec5.1.1}

\begin{figure}[t]
\hspace{-1cm}  \includegraphics[scale=0.57]{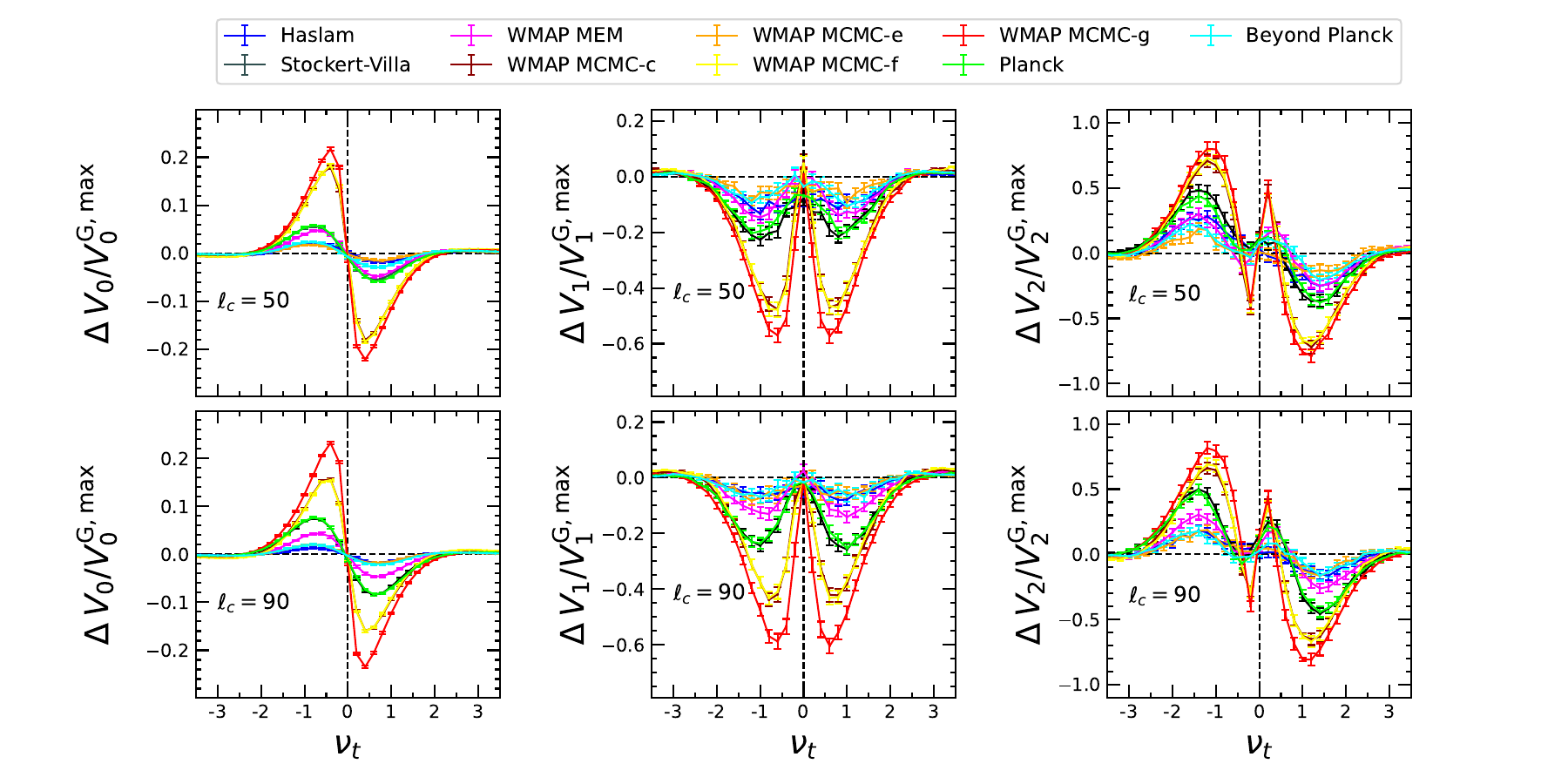}
 \caption{Non-Gaussian deviations $\Delta V_k/V_{k}^{\rm G,max}$ for component separated synchrotron temperature maps.  Error bars denote 2$\sigma$ deviation from the mean values.}
 \label{fig:residnG}
\end{figure}
MFs are powerful tools for detecting the presence of non-Gaussianity whose nature is a priori unknown. In our previous paper~\cite{Rahman:2021azv}), we showed that towards smaller scales, the \has~map is kurtosis-dominated, and skewness is relatively small. This cannot be efficiently detected by the bispectrum but will show up in the trispectrum. Here, we carry out the same analysis to quantify the level and type of non-Gaussianity for the component separated synchrotron maps. In order to distinguish different kinds of non-Gaussianity, it is useful to probe the individual generalized skewness and kurtosis moments (see section 3.2. of \mg{Rahman et al.}~\cite{Rahman:2021azv}). Here we will not show the individual moments but will focus on their consolidated effect.

For each component separated map that we study, we need to calculate $V_k^G$. This is obtained from Gaussian simulations (section \ref{sec:sec2.3.1}) to which instrument noise maps are added. Then we subtract $V_k^G$ from $V_k$ to  obtain  the non-Gaussian deviation $\Delta V_k$. 
For $\ell_c=0$, the fields have $\Delta V_k/V_k^G > 1$, and therefore it is not meaningful to discuss mild deviations from Gaussian nature in terms of perturbative expansions. Therefore, we  focus on smaller scales.
Figure \ref{fig:residnG} shows $\Delta V_{k}/V_k^G$ versus $\nu_t$ for all the component separated maps, for $\ell_c=50$ and 90. The error bars are 2$\sigma$. We make the following observations.

\vskip .1cm
\noindent{\bf Shape of deviations and nature of non-Gaussianity}: the non-Gaussian deviation shapes for all the maps are roughly similar to that of  \has~map. The shapes are characteristic of kurtosis-type non-Gaussianity~\cite{Rahman:2021azv,Matsubara:2011}, with the relative differences depending on which of the generalized kurtosis variables are dominant. Therefore, we infer that the nature of the non-Gaussianity of the fluctuations in the \planck, \SV and \wmap~synchrotron maps are also of kurtosis-type. 

\vskip .15cm
\noindent{\bf Level of non-Gaussian deviations}: for both values of $\ell_c$, we find that \BP~ and \wmap~ \texttt{MCMC-e} have comparable levels of non-Gaussianity with \has, while the other maps show higher levels.  
The level of non-Gaussianity for \planck, \wmap~ \texttt{MCMC-c, f, g} and  \texttt{MEM}, and \SV maps do not show a decrease towards smaller scales (an increase of $\ell_c$). This can be explained by point sources contamination for \planck~ as well as \wmap~ maps, as discussed in the previous subsection.  In addition the \wmap~maps contain the effect of residual contamination by other foregrounds, particularly AME  \cite{Planck2015XXV}.  
A summary of the significance of the non-Gaussian deviations is given  along with the results for polarization in table~\ref{tab:t1} in section \ref{sec:sec5.2}.

\subsubsection{Statistical isotropy}
\label{sec:sec5.1.2} 

\begin{figure}[t]
\begin{center}
\includegraphics[scale=0.58]{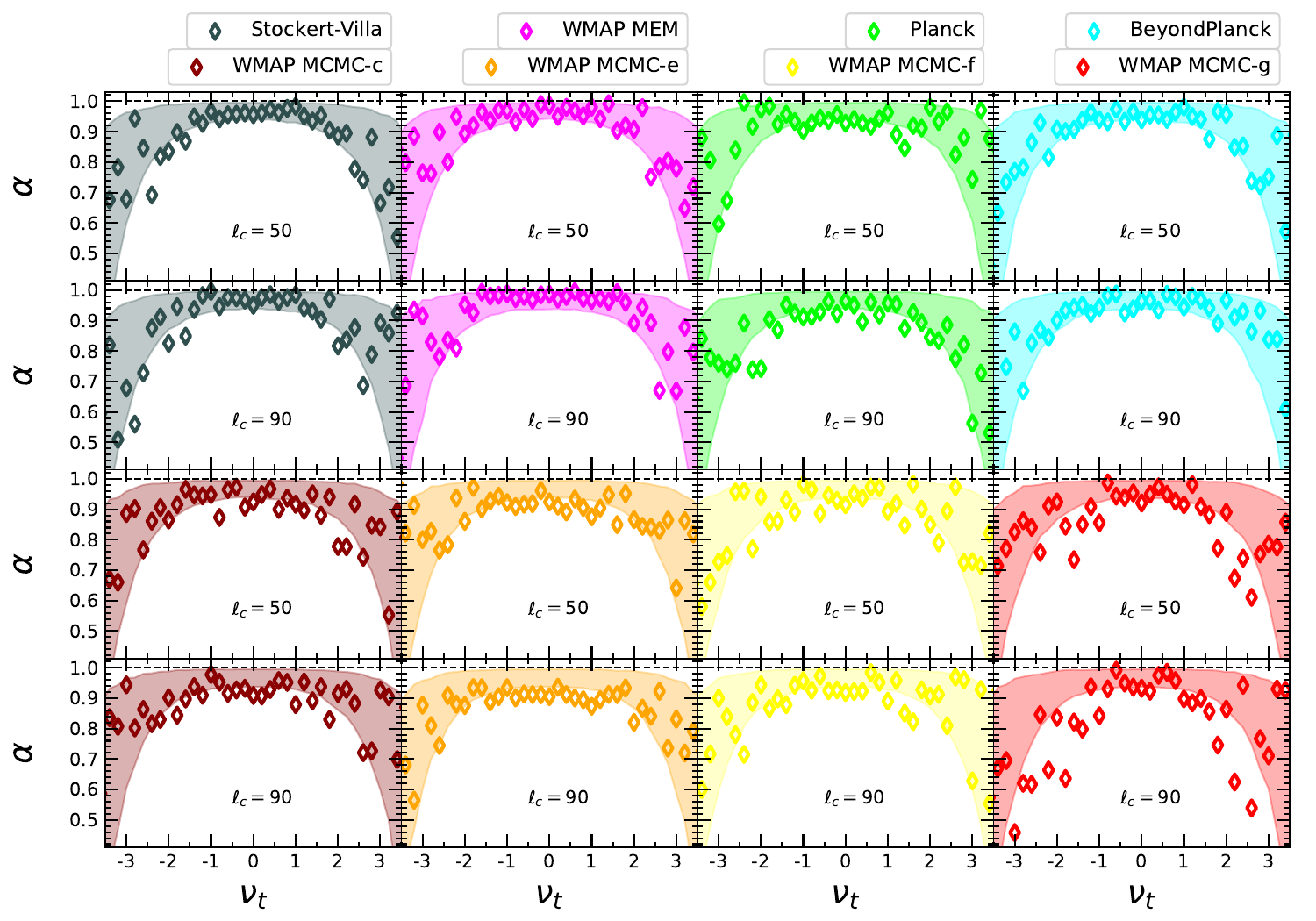} 
\end{center}
\vspace{-0.6cm}
 \caption{Statistical isotropy parameter $\alpha$ versus $\nu_{t}$, for component separated synchrotron maps. The  shaded regions show the 95\% confidence interval about the median values at each threshold obtained from 1000 Gaussian simulations.} 
\label{fig:alpha}
\end{figure}

To probe the statistical isotropy of the component separated synchrotron temperature  maps, we use the $\alpha$ statistic defined in eq.~(\ref{eq:alphadef}). We compute $\alpha$ for all the maps for different bandpass filter scales. We also compute $\a$ for the 1000 Gaussian simulations plus noise corresponding to each map.  
The results for $\alpha$ are shown in figure \ref{fig:alpha} for $\ell_c=50$ and $\ell_c=90$. Each panel corresponds to one map for each $\ell_c$.  The coloured diamonds show $\a$ versus $\nu_t$ for each map. Since $\a$ follows Beta probability distribution,  we compare $\alpha$ of each map with the median values  obtained from the corresponding 1000 Gaussian simulations. The shaded regions correspond to the 95\% confidence levels (CL) about the median values.

Note that values of $\a$ less than the 95\% CL lower limit are the ones with significant anisotropy.  We find that \BP~ and \wmap~\mem~are comparatively the most isotropic for both $\ell_c$ values, with almost all $\alpha$ lying within 95\% CL. \planck~ map shows anisotropy at a few thresholds, and there are more anisotropic threshold values for $\ell_c=90$ compared to 50. This correlates with the higher level of non-Gaussianity towards smaller scales seen earlier, and is likely due to residual point sources in the map. \wmap~\mcmc~\texttt{c, f, g} maps also exhibit anisotropy at several threshold values, and it is visually clear that \texttt{WMAP MCMC-g} shows the highest level of anisotropy at both the scales. Again this correlates with what we found for the level of non-Gaussianity in the previous subsection. We find that \texttt{WMAP MCMC-e} also shows significant anisotropy  at several threshold values for both bandpass filter scales. Since this map shows low level of non-Gaussianity, the significant anisotropy is not in alignment with the other maps for which we find higher anisotropy for higher non-Gaussianity.  

\subsection{Morphology of synchrotron polarization}
\label{sec:sec5.2} 

\begin{figure}[t]
\centering
\includegraphics[height=12cm,width=16cm]{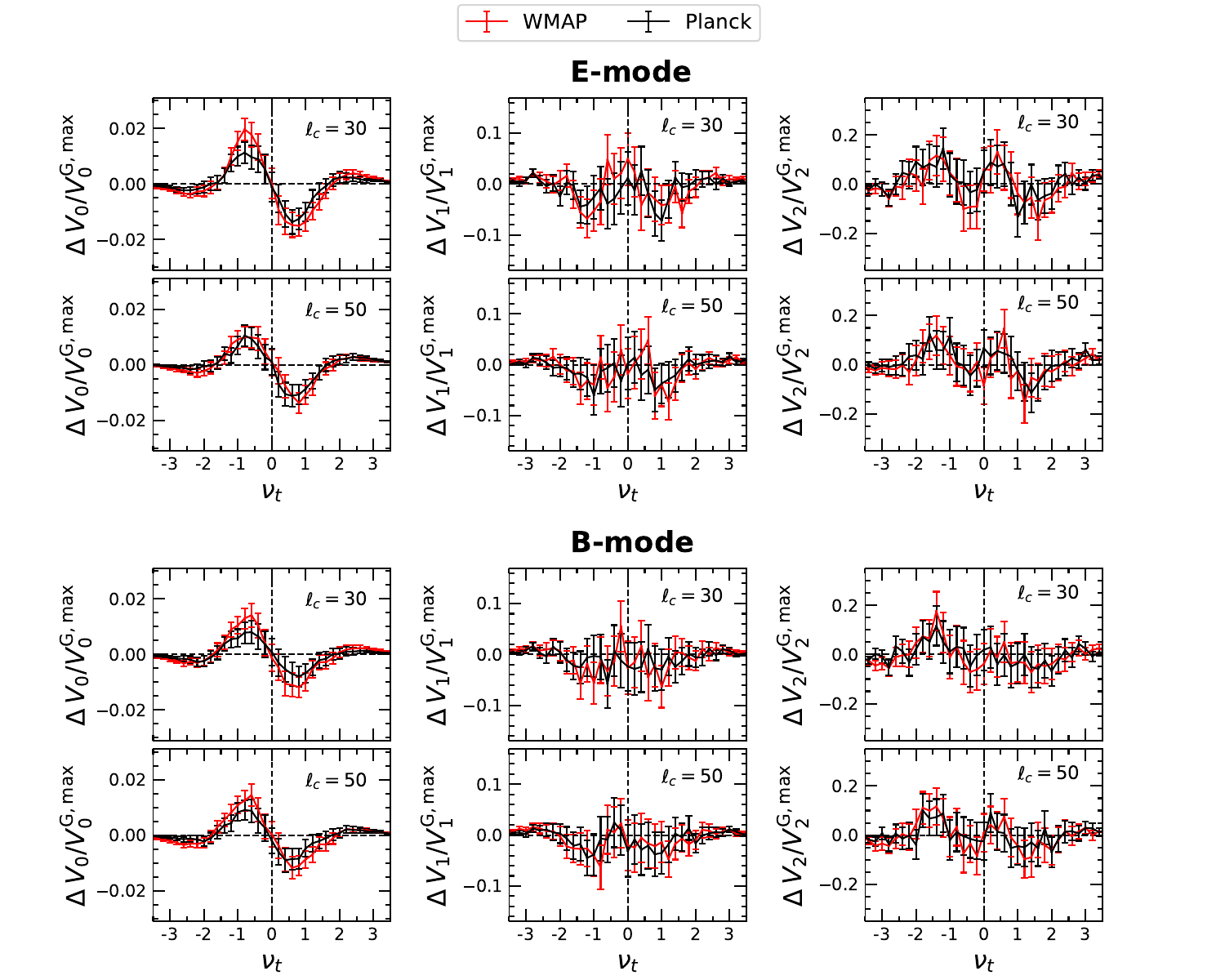}\\
 \caption{Non-Gaussian deviations ($\Delta V_k/V_{k}^{\rm G,max}$)  for component separated $E$ and $B$ mode synchrotron polarization maps from \wmap~and \planck. Error bars are 2$\sigma$ around the mean values. }
\label{fig:dV_EB}
\end{figure}

Next, we discuss the morphology of component-separated synchrotron polarization maps provided by \wmap~and \planck. We focus on bandpass filter scales $\ell_c=30$ and $\ell_c=50$ since instrumental noise dominates in the maps towards smaller scales. We use the \planck~  polarization maps derived using the \texttt{Commander} component separation method. For \wmap, we use the polarization maps derived using the \texttt{MCMC} technique (\texttt{\mg{MCMC-g}}). MFs are calculated for {\it E} and {\it B} mode maps, which are obtained using \texttt{NaMaster}.  
We also compute MFs for 1000 Gaussian simulations generated using the power spectrum of each  map, after adding white noise realizations  (see appendix \ref{sec:a1}). 

Figure~\ref{fig:dV_EB} shows the non-Gaussian deviations of 
the MFs for $E$ and $B$ mode maps of \wmap~and \planck. The shape of the deviations shows excellent agreement with what was obtained for the \has~(temperature) map for both $E$ and $B$ mode maps (compare with figure~\ref{fig:residnG}). This finding confirms that the nature of non-Gaussianity of Galactic synchrotron is of kurtosis-type with the skewness contribution being sub-dominant. 
As was found for the \has~map, we find that $\D V_0$ provides more stringent confirmation of the nature of non-Gaussianity than $\D V_1$ and $\D V_2$.  

\mg{We also conducted our analysis using a different mask, prepared using the \texttt{Planck} synchrotron polarization amplitude ($P=\sqrt{Q^{2}+U^{2}}$) instead of the \texttt{Haslam} map. The mask preparation procedure is outlined in appendix~\ref{sec:a5}. The appendix also shows the polarization amplitude map and the mask used. This mask retains a sky fraction of 61\% for the analysis, which we refer to as {\em P}-mask. Using this mask, MF deviations show similar kurtosis non-Gaussian trends, with slightly different amplitude, suggesting that our findings are robust and not influenced by the choice of the mask. Since the sky fraction for analysis is less (compared to other masks), the non-Gaussian deviations obtained using the $P$-mask have slightly larger error bars. This is reflected in the $\chi^{2}$ calculations discussed in the next subsection.}

\subsection{Statistical significance of non-Gaussian deviations}
\label{sec:sec5.3}
To estimate the significance of non-Gaussian deviations, we calculate $\chi^{2}$ defined as, 
\begin{equation}
\chi^2=\frac{1}{N_{\rm tot}}\sum_{k=1}^{3}\sum_{\nu_{t}=-3}^{3}\frac{\left(V_k^{\rm data}(\nu_{t})-\overline{V}_k^{\rm (G)}(\nu_{t})\right)^{2}}{\sigma^{2}_{V_k^{\rm (G)}}(\nu_{t})}
\end{equation}
$V_k^{\rm data}$ is the MF estimated for different synchrotron maps, $\overline{V}_k^{\rm (G)}(\nu_{t})$ is the mean of the respective Gaussian simulations, and ${\sigma^{2}_{V_k^{\rm (G)}}(\nu_{t})}$ the standard deviation. The threshold bin width is $0.4$. The total number of statistics  is $N_{\rm tot}=16\times 3 = 48$.

In table \ref{tab:t1}, we show the square root of $\chi^2$ values  for different component separated synchrotron temperature and polarization maps. For temperature maps, we find that the values are quite large, indicating that the maps  are highly non-Gaussian, even at the relatively smaller scales considered here. Compared to \has~map, all maps  except \wmap~\mcmc-\texttt{e} \mg{and \BP~}have higher levels of non-Gaussianity that increase as $\ell_c$ increases. \mg{\wmap~\mcmc-\texttt{e} also show a slight increment as we go to small scales, unlike \BP. This again signifies the better treatment of point sources in the \BP~analysis.} As discussed in section~\ref{sec:sec5.2} and also in appendix~\ref{sec:a4}, our analysis strongly suggests that these derived synchrotron products contain residual contamination by other Galactic components, with some contribution from unresolved point sources.
It should be noted that the $\chi^{2}$ values given in \mg{Rahman et al.}~\cite{Rahman:2021azv} were estimated for a sky fraction of 40\%. The current analysis, however,  is  carried out on 65\% of the sky to retain the sky regions with a high signal-to-noise ratio. Hence, they cannot be compared directly. 

For polarization, the $\chi^{2}$ values are smaller than the values  of temperature. All the maps 
show a decrease of $\chi^{2}$ towards smaller angular scales. We find that the main contribution to the $\chi^{2}$ is coming from $V_{0}$ for all the temperature and polarization maps. This is evident from figures \ref{fig:residnG} and \ref{fig:dV_EB}, where $V_{0}$ has error bars smaller than that of $V_{1}$ and $V_{2}$. This was also observed in our previous analysis of \has~map. In the case of polarization, the $\chi^2$ values are lower due to the increased level of noise present in these maps. \mg{The values enclosed in parentheses represent the results obtained using the $P$-mask. They are slightly lower due to the reduced sky coverage in the analysis, resulting in larger error bars.}

We have repeated the analysis with the {\em filament mask} (right panel of figure~\ref{fig:mask_bp}) for both temperature and polarization maps and found no significant difference from the results discussed above. This suggests that large-scale filamentary structures in the synchrotron sky do not affect our results.

\renewcommand{\arraystretch}{1.6}
\begin{table}[t]
\begin{tabular}{|p{0.5cm}|p{1.10cm}|p{1.5cm}|p{1.1cm}|p{1.15cm}|p{1.15cm}|p{1.15cm}|p{1.15cm}|p{1.1cm}|p{1.1cm}|}
\hline
\multicolumn{10}{|c|}{\bf Temperature} \\
\hline
$\ell_c$ & \texttt{Haslam} & \texttt{Stockert- Villa} & \texttt{WMAP} \texttt{MEM} & \texttt{WMAP} \texttt{MCMC-c} & \texttt{WMAP} \texttt{MCMC-e} & \texttt{WMAP} \texttt{MCMC-f} & \texttt{WMAP} \texttt{MCMC-g} & \texttt{Planck} & \texttt{Beyond Planck} \\
\hline
$50$ & 9.54 & 19.49 & 16.18 & 42.19 & 7.15 & 42.20 & 50.60 & 19.83  & 9.75 \\
\hline
$90$ & 7.33 & 29.25 & 17.93 & 46.01 & 8.66 & 46.25 & 63.23 & 29.95 &  9.19 \\
\hline

\multicolumn{10}{|c|}{\bf Polarization} \\
\hline
$\ell_{c}$ & \quad --- & \multicolumn{2}{|c|}{\texttt{WMAP E-mode}}    & \multicolumn{2}{|c|}{\texttt{WMAP B-mode}}   & \multicolumn{2}{|c|}{\texttt{Planck E-mode}}  & \multicolumn{2}{|c|}{\texttt{Planck B-mode}}  \\
\hline
$30$ & \quad --- & \multicolumn{2}{|c|}{{3.65}~\mg{(3.30)}}    & \multicolumn{2}{|c|}{{3.10}~\mg{(2.78)}}   & \multicolumn{2}{|c|}{{2.48}~\mg{(1.85)}}  & \multicolumn{2}{|c|}{{1.94}~\mg{(1.93)}} \\
\hline
$50$ & \quad --- & \multicolumn{2}{|c|}{{2.81}~\mg{(2.91)}}    & \multicolumn{2}{|c|}{{3.26}~\mg{(3.17)}}   & \multicolumn{2}{|c|}{2.71~\mg{(1.70)}}  & \multicolumn{2}{|c|}{2.18~\mg{(1.87)}} \\
\hline
\end{tabular}
\caption{The square root of $\chi^{2}$, quantifying the non-Gaussian deviations of  different component separated synchrotron temperature and polarization maps.}
\label{tab:t1}
\end{table}

\section{Summary and discussion}
\label{sec:sec6}

In this paper our primary goal is to understand the nature of non-Gaussian deviations of Galactic synchrotron emissions on different length scales. For this purpose, we analyzed the morphological properties of two sets of Galactic foreground maps.

The first set of maps we analyze is comprised of observed Galactic total emissions at different frequencies ranging from 408 MHz to 33 GHz. This analysis reveals how the morphology of the total Galactic foreground transforms as synchrotron emission, which is the main component towards lower frequencies, becomes the sub-dominant component at 23 GHz and higher frequencies. From the results of this study, we can anticipate the impact of residual contamination by other foreground components on the morphology of component separated synchrotron maps. Since the observed frequency maps have high SNR, it is unlikely that our results are biased by instrumental noise. \mg{Furthermore, we have demonstrated that the results we have obtained remain unaffected by the instrumental systematics inherent in these maps.} We then compare them with the morphology of simulated total foreground emission maps obtained from PySM. This comparison reveals significant amplitude and shape differences of the MFs between the observed maps and the simulations on large scales and relatively better agreement towards smaller scales.

The second set of maps we analyze are component separated  synchrotron temperature maps from \wmap, \planck~and \texttt{BeyondPlanck}, and polarization maps from \wmap~and \planck. From all the maps studied,  both temperature and polarization,  we conclude that the nature of non-Gaussian deviation of small-scale fluctuations of synchrotron emission is of kurtosis type. This is in agreement with our earlier finding from \has~map~\cite{Rahman:2021azv}. We have included instrumental noise in estimating the uncertainties. \mg{Also, we have taken into account the role of instrumental systematics in our results, and it has been found to be minimal.} Hence our conclusion is not biased by such effects.
This result is important from two perspectives. First, it provides a concrete direction for modelling small-scale fluctuations of synchrotron emission as mildly non-Gaussian fields of kurtosis nature, rather than Gaussian. The isotropic nature of \BP~map, in agreement with \has~map \cite{Rahman:2021azv}, also implies that the modelling of small-scale fluctuations of synchrotron as a statistically isotropic field is in the right direction. These findings can therefore  improve component separation pipelines, which is particularly important for B-mode experiments. In fact, Martire et al.~\cite{Martire:2023ytg} have recently implemented an algorithm for including the kurtosis nature in generating non-Gaussian synchrotron models. Secondly, this result implies that contamination of the true CMB by residual synchrotron component will most likely result in kurtosis-type non-Gaussianity. Hence, it will not be detectable by 3-point statistics such as the bispectrum. Preliminary work to determine the type of non-Gaussianity induced by residual foregrounds in \wmap~data using MFs was carried out in \mg{Chingangbam \& Park}~\cite{Chingangbam:2013}. It is timely to revisit such analysis in the light of our current results. 

Lastly, we comment on the comparison of the morphology of different component separated synchrotron temperature maps. This comparison serves to analyze the performance of various pipelines that are adopted for \wmap~and \planck~data, beyond the usual zeroth order comparison at the map and power spectrum level.  Accurate component separation by each independent method must lead to the same morphology, that is, a similar level of non-Gaussianity and statistical isotropy.  
Using \has~map as the benchmark,  we find that  \texttt{BeyondPlanck} and \wmap~\texttt{MCMC-e} are the best-performing pipelines. The other maps show differences that are significant. In particular, \planck~map shows a significant difference towards small scales, which is likely to be due to the presence of unresolved point sources. The other \wmap~maps show even larger differences at different scales. This is likely to be due to unresolved point sources, and also contamination by residual AME and/or free-free emissions. A systematic investigation of the various contributions from residual contamination and point sources in the future will be very valuable.

Our results  underscore the need for further improvement of the  component separation techniques. It will be interesting to extend  our analysis to other low-frequency surveys like S-PASS (at 2.3 GHz)~\cite{Carretti2019}, C-BASS (at 5 GHz)~\cite{Stuart2022}, and the QUIJOTE-MFI (from 11 to 19 GHz)~\cite{QUIJOTEIV:2023}. We would  like to explore the morphological properties of other Galactic components and plan to carry out these extensions in the near future. It is also of importance to provide a physical explanation for the kurtosis nature of Galactic synchrotron emissions.

\acknowledgments 
    {We thank the anonymous referee whose insightful comments and constructive feedback enhanced the quality of this manuscript. We acknowledge the use of the \texttt{Nova} cluster at the Indian Institute of  Astrophysics, Bangalore. 
    We have used \texttt{HEALPIX}\footnote{\url{http://healpix.sourceforge.net/}}~\cite{Gorski:2005}, \texttt{healpy}~\cite{,Zonca:healpy}, \texttt{Matplotlib}~\cite{Hunter:2007}, PySM~\cite{Thorne2017}, \texttt{NaMaster}~\cite{NaMaster:2019} packages for producing the results in this paper.  We thank M. Vidal and C. Dickinson for providing the filament mask prepared by them. F. R. would like to thank H. K. Eriksen, D. J. Watts, other members of the Oslo CMB group, S. E. Harper and T. J. Rennie  for useful discussions. F. R. acknowledges the support of the Korea Institute for Advanced Study, Seoul, for the academic visit where a part of this work was carried out.}


\appendix

\section{Estimation of instrumental noise }
\label{sec:a1}

The uncertainties of the MFs for the observed frequency maps and the component separated synchrotron maps shown in the paper include instrumental noise. For estimating the uncertainties we need noise maps for each experimental setup. 
For \planck~LFI 30 GHz frequency map, we do not produce noise simulations ourselves. We use 300 FFP10 noise simulations provided by \planck~\cite{Planck_2018III_HFI}. For the other observed frequency and component separated maps the instrumental noise properties  differ from map to map. We generate them using noise characteristics provided by the experiment.  We describe them case by case below. 
\begin{itemize}

   \item For \has~and \SV maps, we are provided with the noise rms values of the instruments. These are 800 mK~\cite{Remazeilles:2014mba} and 17 mK~\cite{testori_radio_2001},  respectively. For each observed map, we generate 1000 noise maps, each of which are Gaussian random numbers at each pixel with the respective rms values as the standard deviation. 

    \item To generate the noise maps for \wmap~K and Ka frequency maps, we use the following equation
      \begin{equation}\label{eq:noise_eq}
        \sigma^{2}(\hat{n})=\sigma_{0}^2/N_{\rm obs}(\hat{n}),
    \end{equation}
where $\hat n$ is the sky direction, $\sigma_0$ is the noise rms  and  ${N}_{\rm obs}$ is the number of observations taken at each pixel. For K band, $\sigma_{0}=1.429\, {\rm mK}$, and for Ka band it is  $1.466\, {\rm mK}$~\cite{bennett_nine-year_2013}.  We  use ${N}_{obs}$ for K and Ka bands provided by \wmap. 
Then we generate Gaussian random numbers at each pixel with $\s(\hat n)$ as the standard deviation. We generate 1000 such noise maps.

   \item  The component separated \wmap~\mcmc, \planck~and \BP~data sets include the synchrotron posterior rms maps. We use these map values as the standard deviation for generating 1000 Gaussian random numbers at every pixel.

    \item For \wmap~\mem~synchrotron map, noise rms is not provided. So, we identify the most noisy frequency channel, which is W2, to generate the noise maps. This makes our error estimates  conservative. We use $\sigma_{0}=6.94\, {\rm mK}$ ~\cite{bennett_nine-year_2013} and ${N}_{\rm obs}$ map provided by \wmap. We then generate the noise map using eq.~(\ref{eq:noise_eq}).
   
    This map is then used as the noise rms map for generating 1000 Gaussian random numbers at each pixel. 

    \item For \wmap~\mcmc~polarization maps, the $Q,U$ posterior rms maps are provided. These map values serve as the standard deviation for generating noise realizations, similar to what is followed for temperature maps. 
    
    \item For \planck~\texttt{Commander} polarization maps, we consider the difference between two half-mission maps as the noise realization. 
    
 \end{itemize}

\section{Signal-to-noise ratio of frequency maps of WMAP and Planck}
\label{sec:a2}

\begin{figure}[t]
\centering
\includegraphics[scale=0.36]{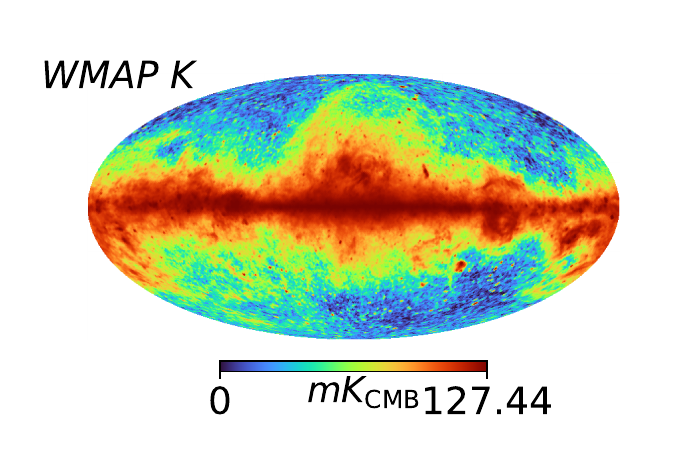} \quad \quad
\includegraphics[scale=0.36]{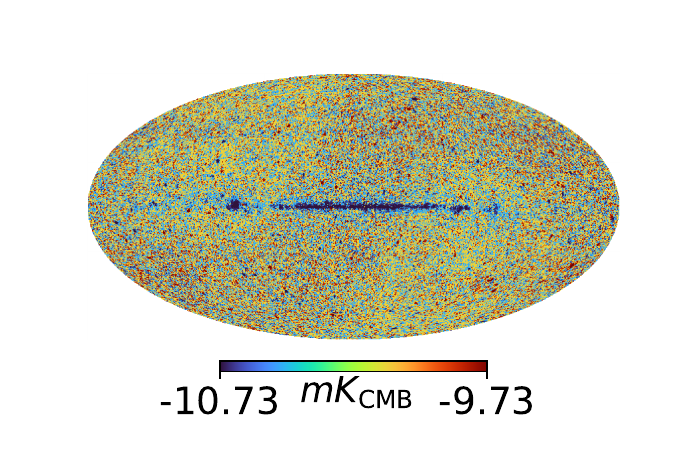}  \quad \quad
\includegraphics[scale=0.36]{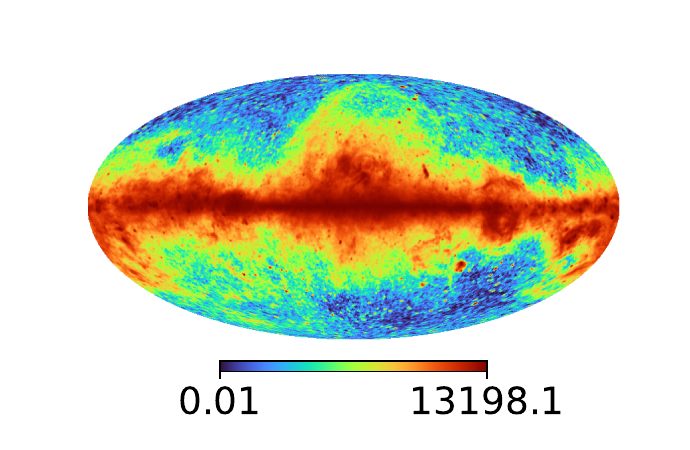} \\
\vspace{3mm}
\includegraphics[scale=0.36]{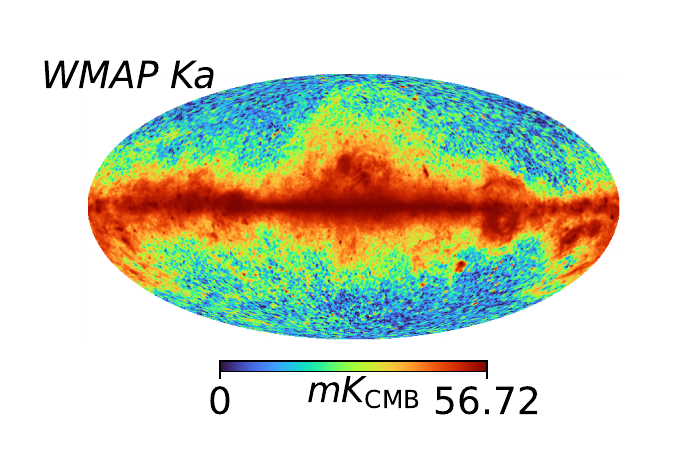} \quad \quad  
\includegraphics[scale=0.36]{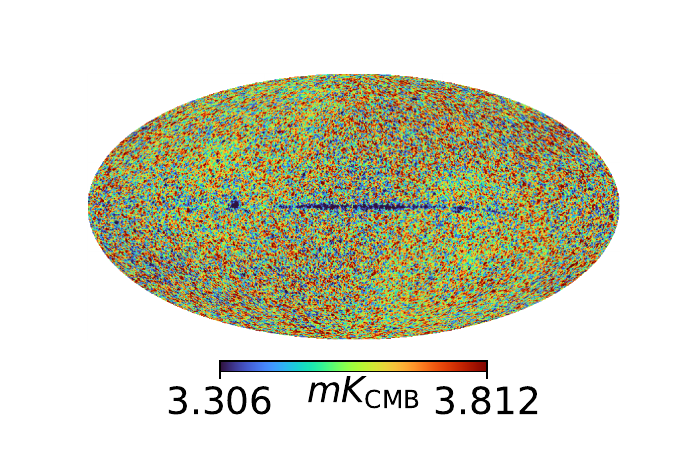}\quad \quad 
\includegraphics[scale=0.36]{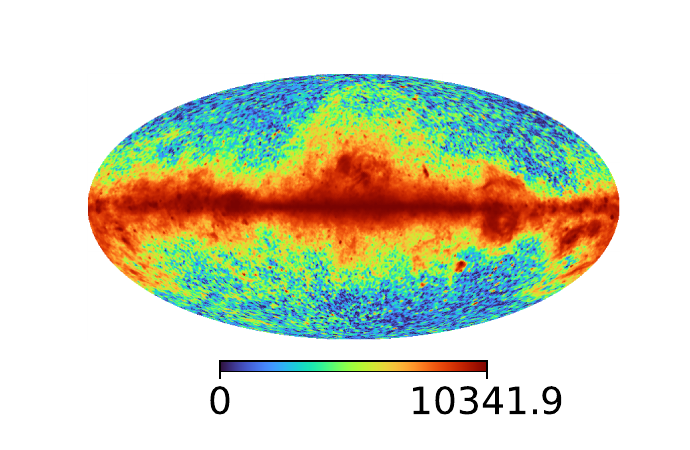}
\\
\vspace{3mm}
\includegraphics[scale=0.36]{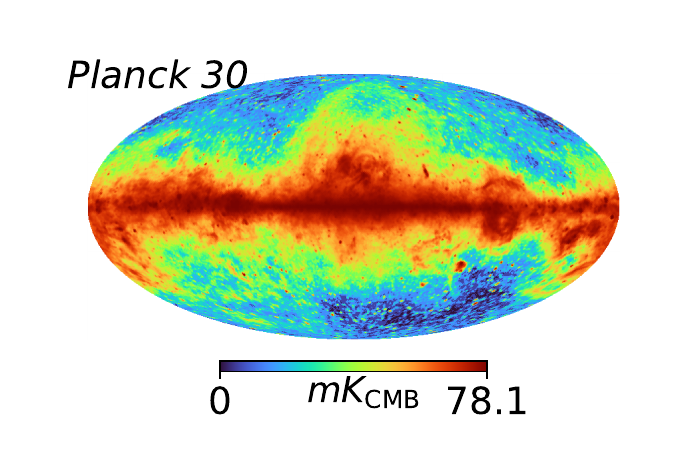} \quad \quad 
\includegraphics[scale=0.36]{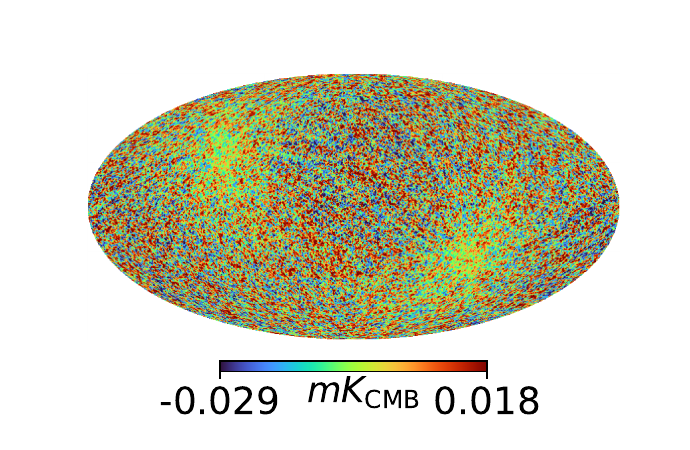} \quad \quad 
\includegraphics[scale=0.36]{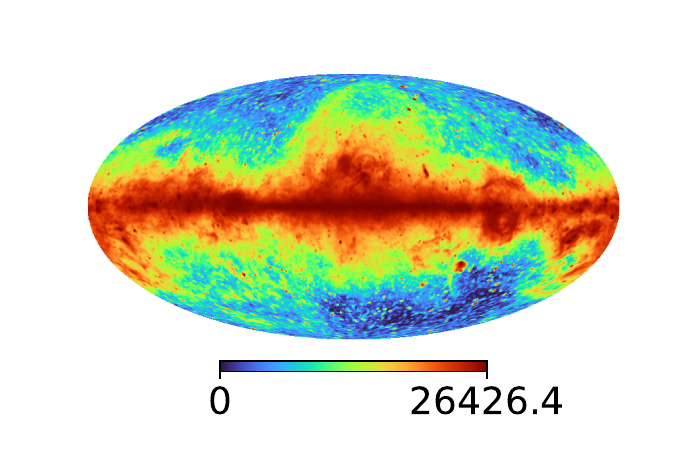} 
\\

 \caption{{\it Left column:} Maps of \wmap~K band (top), Ka band (middle) and \planck~LFI 30 GHz (bottom), after subtracting the best-fit CMB map given by the respective experiments. {\it Middle column:} Maps showing the instrumental noise in the corresponding map on the left. {\it Right column}: Maps of the SNR for the same.}
\label{fig:snr_maps}
\end{figure}

In this section, 
we estimate the signal-to-noise ratio (SNR) of \planck~LFI 30 GHz map, \wmap~K band and Ka band maps. 
The left panel of figure \ref{fig:snr_maps} shows \wmap~K and Ka and \planck~LFI 30 frequency maps after subtracting the best fit CMB maps from \wmap~and \planck, respectively. The middle panel shows the maps containing instrumental noise from the respective maps in the left panel. These noise maps are obtained as explained below:
\begin{itemize}
    \item For \planck, the noise map is generated by subtracting the two half-ring maps. Half-ring maps are the maps prepared using only the first or second halves of each ring (pointing period) of the telescope, so  the sky signal is expected to be the same, but the noise will differ. By subtracting these maps, we can isolate the noise content.  
    \item For \wmap~K and Ka bands, we average the observed maps of the first four individual years, and similarly average for the next 4 years. Then we subtract the two resulting maps  to estimate the noise. 
\end{itemize}

We then obtain the signal-to-noise ratio (SNR) map which is the signal map (shown in the left panel) divided by the standard deviation of the corresponding noise map (shown in the middle panel). This map is shown in the third column of figure \ref{fig:snr_maps}.  We find that the SNR maps of both \wmap~and \planck~have high values across all the pixels.

\section{\mg{Effect of instrumental systematics}}
\label{sec:a3}

\begin{figure}[t]
\begin{center}
\includegraphics[scale=0.56]{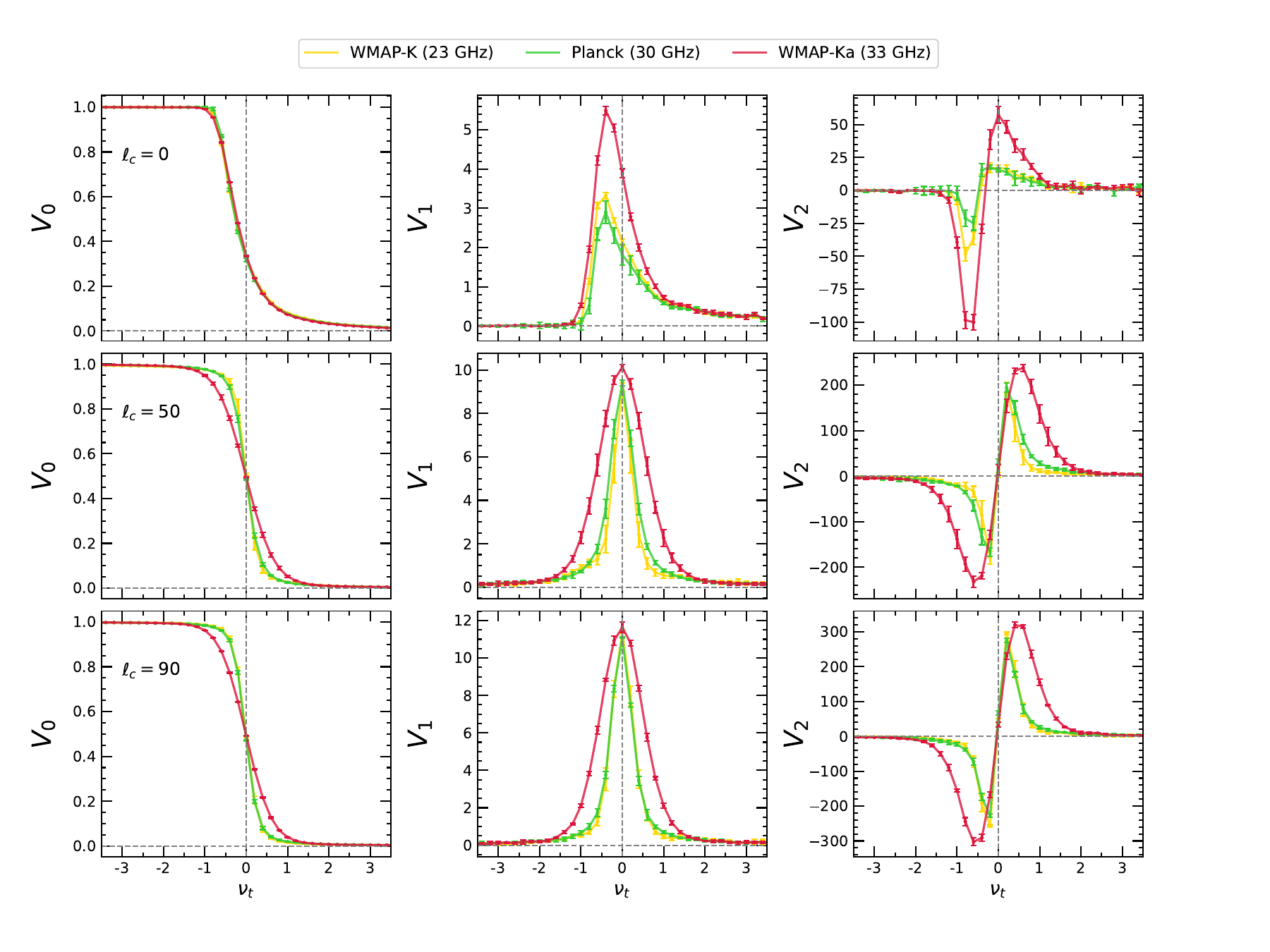}
\vspace{-11 mm}
\caption{{\em Top}: Mean and 2$\sigma$ error bars of MFs for \texttt{WMAP} and \texttt{Planck} individual year frequency maps at small angular scales.}
\label{fig:syst_plot}
\end{center}
\end{figure}

\mg{Instrumental effects can introduce alterations to the morphology of the maps we have been studying, potentially affecting our overall analysis and interpretation. This section explores the possible role of instrumental systematics in the morphological distinctions we have observed in the maps we studied. Our focus lies on \texttt{WMAP} and \texttt{Planck} products, as examining the instrumental effects of \texttt{Stockert-Villa} requires a comprehensive understanding of the instrument and calibration techniques. We reserve this topic for a future study.}

\mg{To quantify this, we focus on the observed frequency maps of \texttt{WMAP} (K and Ka) and \texttt{Planck} (30 GHz), which largely contribute to the derived synchrotron products of the respective missions. We use the individual year maps at each frequency (9 years for \texttt{WMAP} and 4 years for \texttt{Planck}) and study how the morphology varies at each year. Under the assumption that the signal remains consistent across each year, while the noise and instrumental effects vary with time, we obtain an estimate of how significant the noise and systematics are, in each frequency bands. We calculate the three scalar MFs for these maps at different angular scales ($\ell_{c}=0$, $\ell_{c}=50$ and $\ell_{c}=90$). The results of the mean MFs and 2$\sigma$ error bars are shown in Figure~\ref{fig:syst_plot}. The error bars represent the preliminary assessment of both the systematics and noise present in each map.}

\mg{We find that the mean MFs for each frequency map fall beyond the 2$\sigma$ error bars of other maps. This indicates that the systematics has negligible effect in the observed morphological distinctions of the frequency maps. To put it differently, the morphological differences we are observing are significantly larger than the systematic uncertainties. This trend is seen for all the three MFs at all the scales we studied. Consequently, the synchrotron products derived using these maps remain unaffected by the instrumental effects. Therefore, our analysis implies that the morphological distinctions we observe for different synchrotron maps at both small and large scales are not biased by the instrumental systematics.}\\ 

\section{Morphology of composite foreground fields}
\label{sec:a4}

\begin{figure}[t]
\begin{center}
\includegraphics[scale=0.53]{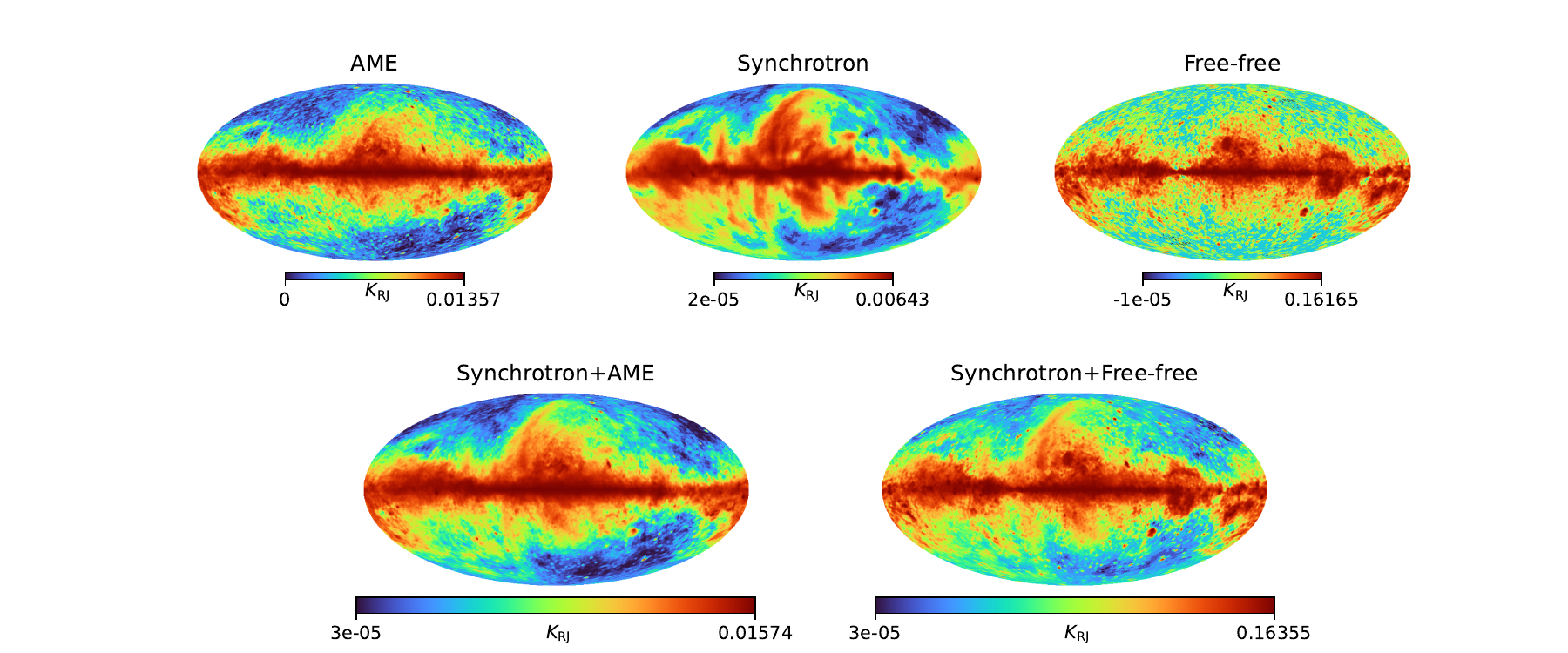}\\
\caption{{\em Top}: Maps of AME (left), synchrotron (middle) and free-free (right) emissions at 23 GHz generated using PySM. {\em Bottom}: Maps of the sums of synchrotron and AME (left), and synchrotron and free-free (right).}
\label{fig:afs}
\end{center}
\end{figure}

Here, we want to probe how the MFs of the sum of two foreground components differ from the MFs of the individual components. For this purpose, we generate maps of synchrotron, AME and free-free emissions at 23 GHz using PySM. The maps and their sums are shown in figure~\ref{fig:afs}. Note that the ranges of the field values shown differ from map to map. At 23 GHz, free-free is the most dominant, followed by AME and synchrotron. Visually, we can also see that synchrotron fluctuations are considerably smoother (implies less power on smaller scales) compared to AME and free-free. 

\begin{figure}[t]
\begin{center}
\vspace{1mm}
\includegraphics[scale=0.42]{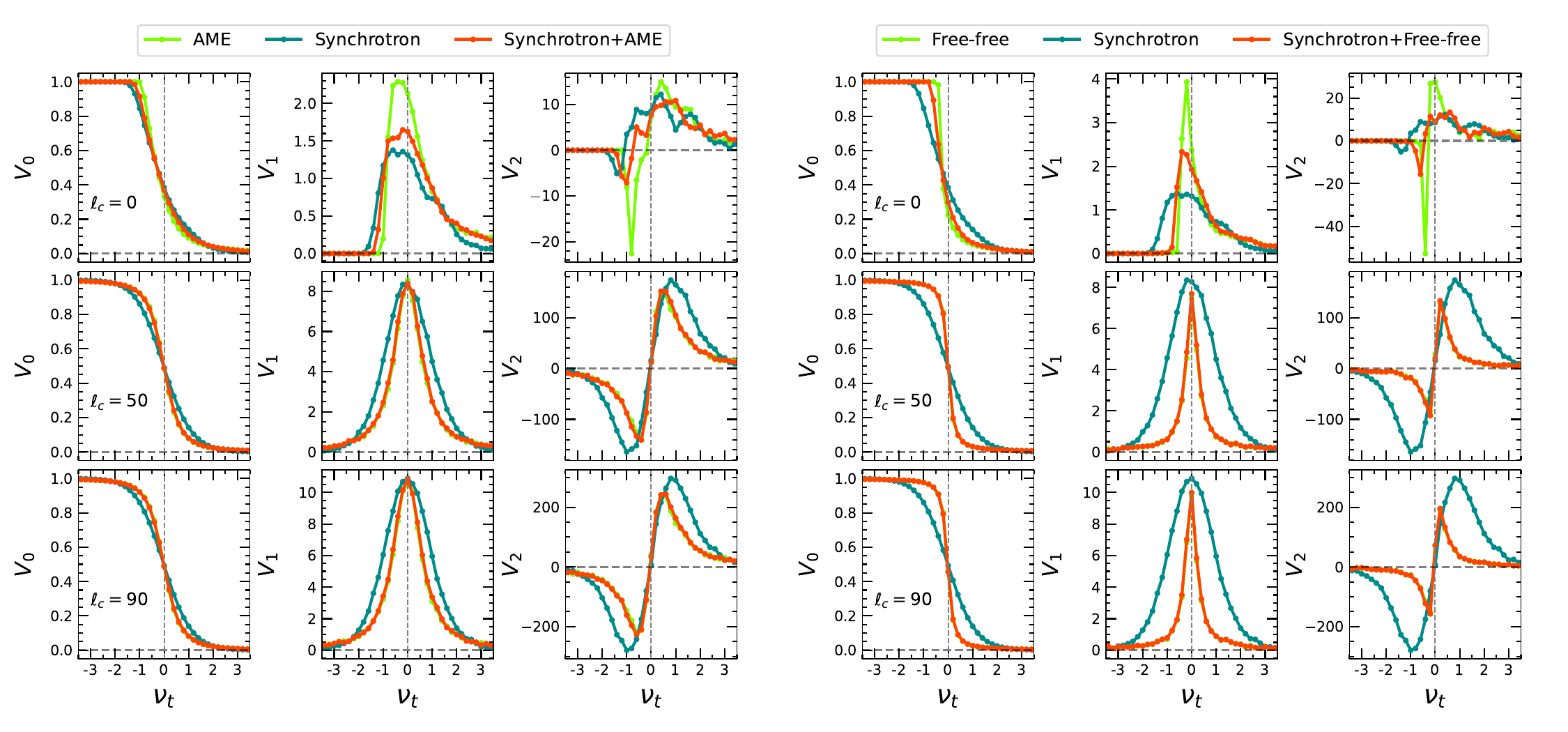}    
\end{center}
\vspace{-0.6cm}
\caption{MFs for the foreground component maps shown in figure~\ref{fig:afs}.}
\label{fig:afs_mfs}
\end{figure}

For each of the five maps, we compute MFs after bandpass filtering and masking identically.  The results are shown in figure~\ref{fig:afs_mfs}.  The first three columns towards the left compare the MFs of AME, synchrotron and their sum in each panel, with the rows from top to bottom corresponding to $\ell_c=0,50$ and 90. For $\ell_c=0$ we can clearly see that $V_1$ and $V_2$ for AME have higher amplitudes compared to synchrotron. This can be  understood by comparing their typical size of structures quantified by $\theta_c=\s_0/\s_1$. The amplitudes of $V_1$ and $V_2$ are set by $\theta_c^{-1}$ and $\theta_c^{-2}$, respectively, at the zeroth order Gaussian approximation of any given field.  Synchrotron has larger $\theta_c$ compared to AME due to its relatively smoother nature,  and as a consequence we can expect, and we find, lower amplitudes for $V_1$ and $V_2$.  For $\ell_c=50$ and 90, we find that the MFs of the sum of the two fields are determined primarily by that of AME (overlapping red and green plots). This is due to AME being the dominant field at 23 GHz as well as the relative smoothness of synchrotron.

The last three columns towards the right of figure~\ref{fig:afs_mfs} compare the MFs of free-free, synchrotron and their sum, similar to the comparison with AME. Again synchrotron is much smoother than free-free. We again obtain the MFs of the sum of the two fields to be roughly average of the MFs of the individual fields for $\ell_c=0$, while for $\ell_c=50$ and 90, the morphology of the sum is primarily determined by free-free. 

It is straightforward to extend the above comparison to the sum of the three fields. For the frequency used here, 23 GHz, the morphology of the sum will be determined mainly by free-free since it is the dominant component. We stress that the results will vary with observing frequency as different component fields dominate at different frequencies. We have not considered thermal dust emission here since it remains sub-dominant at the frequencies considered in our paper.  A point that is interesting to note is that AME and free-free emissions are relatively more non-Gaussian than synchrotron at $\ell_c=90$, as can be discerned from the shapes of $V_1$ and $V_2$ and comparison with eq.~(\ref{eqn:GMF}). This shows that in constructing models for these fields the `small' scale beyond which they can be assumed to be Gaussian if they approach Gaussian behaviour at all will be smaller than that of synchrotron. We also note that the use of these findings to interpret the results in section~\ref{sec:sec4.2} is based on the assumption that the models of the emissions input in PySM are accurate. If PySM is inaccurate there will be some variation of the morphologies. However, the inferences for composite fields will broadly remain valid. 

\section{\mg{Analysis with polarization mask}}
\label{sec:a5}

\mg{We discuss the preparation of the polarization mask ($P$-mask) used for the polarization analysis. The method followed is the same as the mask preparation steps outlined in section~\ref{sec:sec3.2}, except that the high emission pixels are identified using the synchrotron polarization amplitude map ($P=\sqrt{Q^{2}+U^{2}}$), instead of the \texttt{Haslam} temperature map. For this, we use the polarization maps provided by \texttt{Planck} using the \texttt{Commander} method.}

\begin{figure}[t]
\begin{center}
\vspace{1mm}
\includegraphics[scale=0.6]{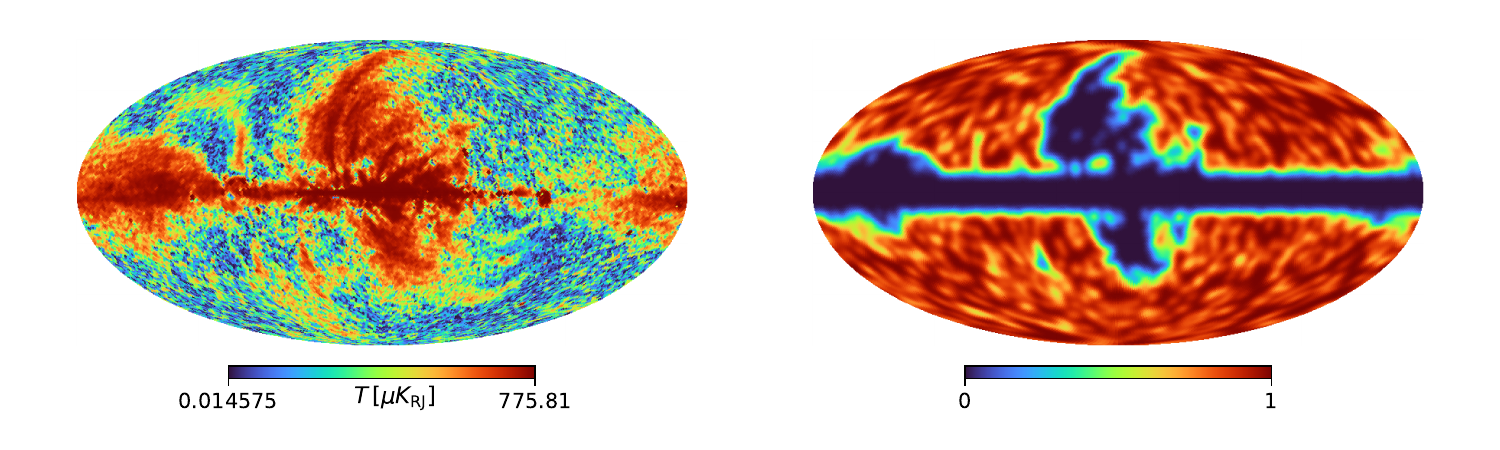}    
\end{center}
\vspace{-0.6cm}
\caption{{\em Left:} Synchrotron polarization amplitude map given by \texttt{Planck}.  {\em Right:} Polarization mask ($P$-mask) prepared based on the polarization amplitude map. The mask is apodized with a Gaussian beam of ${\rm FWHM}=300$ arcmin and has a sky-fraction of 61\%.}
\label{fig:Pmap_Pmask}
\end{figure}

\mg{Figure~\ref{fig:Pmap_Pmask} shows the \texttt{Planck} polarization amplitude map (left panel) and the polarization mask prepared (right). The mask retains a sky-fraction of 61\% for the analysis. For polarization, we work with $N_{\rm side}=64$, and, hence, the mask is apodized with a Gaussian beam of ${\rm FWHM}=300$ arcmin to avoid the sharp boundary effects.}

\def\apj{ApJ}%
\def\mnras{MNRAS}%
\def\aap{A\&A}%
\def\apjl{ApJ}
\def\aj{AJ}
\def\physrep{PhR}
\def\apjs{ApJS}
\def\jcap{JCAP}
\def\pasa{PASA}
\def\pasj{PASJ}
\def\nat{Natur}
\def\apss{Ap\&SS}
\def\araa{ARA\&A}
\def\aaps{A\&AS}
\def\ssr{Space Sci. Rev.}
\def\pasp{PASP}
\def\na{New A}
\def\prd{PRD}

\bibliographystyle{JHEP}
\bibliography{syncref}

\newcommand{\noop}[1]{}

\providecommand{\href}[2]{#2}\begingroup\raggedright\begin{thebibliography}{10}

\bibitem{Rahman:2021azv}
F.~Rahman, P.~Chingangbam and T.~Ghosh, \emph{{The nature of non-Gaussianity and statistical isotropy of the 408 MHz Haslam synchrotron map}}, \href{https://doi.org/10.1088/1475-7516/2021/07/026}{\emph{JCAP} {\bfseries 07} (2021) 026} [\href{https://arxiv.org/abs/2104.00419}{{\ttfamily 2104.00419}}].

\bibitem{Eriksen:2004ss}
H.K.~{Eriksen}, I.J.~{O'Dwyer}, J.B.~{Jewell}, B.D.~{Wandelt}, D.L.~{Larson}, K.M.~{G{\'o}rski} et~al., \emph{{Power Spectrum Estimation from High-Resolution Maps by Gibbs Sampling}}, \href{https://doi.org/10.1086/425219}{\emph{Astrophys. J., Suppl. Ser.} {\bfseries 155} (2004) 227} [\href{https://arxiv.org/abs/astro-ph/0407028}{{\ttfamily astro-ph/0407028}}].

\bibitem{Delabrouille:2008qd}
J.~{Delabrouille}, J.F.~{Cardoso}, M.~{Le Jeune}, M.~{Betoule}, G.~{Fay} and F.~{Guilloux}, \emph{{A full sky, low foreground, high resolution CMB map from WMAP}}, \href{https://doi.org/10.1051/0004-6361:200810514}{\emph{Astron. Astrophys.} {\bfseries 493} (2009) 835} [\href{https://arxiv.org/abs/0807.0773}{{\ttfamily 0807.0773}}].

\bibitem{Tegmark:1999ke}
M.~{Tegmark}, D.J.~{Eisenstein}, W.~{Hu} and A.~{de Oliveira-Costa}, \emph{{Foregrounds and Forecasts for the Cosmic Microwave Background}}, \href{https://doi.org/10.1086/308348}{\emph{Astrophys. J.} {\bfseries 530} (2000) 133} [\href{https://arxiv.org/abs/astro-ph/9905257}{{\ttfamily astro-ph/9905257}}].

\bibitem{Jelic:2008jg}
V.~{Jeli{\'c}}, S.~{Zaroubi}, P.~{Labropoulos}, R.M.~{Thomas}, G.~{Bernardi}, M.A.~{Brentjens} et~al., \emph{{Foreground simulations for the LOFAR-epoch of reionization experiment}}, \href{https://doi.org/10.1111/j.1365-2966.2008.13634.x}{\emph{Mon. Notices Royal Astron. Soc.} {\bfseries 389} (2008) 1319} [\href{https://arxiv.org/abs/0804.1130}{{\ttfamily 0804.1130}}].

\bibitem{Marthi:2017}
V.R.~{Marthi}, S.~{Chatterjee}, J.N.~{Chengalur} and S.~{Bharadwaj}, \emph{{Simulated predictions for H I at z = 3.35 with the Ooty Wide Field Array - I. Instrument and the foregrounds}}, \href{https://doi.org/10.1093/mnras/stx1796}{\emph{\mnras} {\bfseries 471} (2017) 3112}.

\bibitem{Rybicki1986}
G.B.~{Rybicki} and A.P.~{Lightman}, \emph{{Radiative Processes in Astrophysics}} (1986).

\bibitem{kogut2012}
A.~Kogut, \emph{Synchrotron spectral curvature from 22 mhz to 23 ghz}, \href{https://doi.org/10.1088/0004-637X/753/2/110}{\emph{The Astrophysical Journal} {\bfseries 753} (2012) 110}.

\bibitem{Jung2018}
G.~{Jung}, B.~{Racine} and B.~{van Tent}, \emph{{The bispectra of galactic CMB foregrounds and their impact on primordial non-Gaussianity estimation}}, \href{https://doi.org/10.1088/1475-7516/2018/11/047}{\emph{\jcap} {\bfseries 2018} (2018) 047} [\href{https://arxiv.org/abs/1810.01727}{{\ttfamily 1810.01727}}].

\bibitem{Coulton2019}
W.R.~{Coulton} and D.N.~{Spergel}, \emph{{The bispectrum of polarized galactic foregrounds}}, \href{https://doi.org/10.1088/1475-7516/2019/10/056}{\emph{\jcap} {\bfseries 2019} (2019) 056} [\href{https://arxiv.org/abs/1901.04515}{{\ttfamily 1901.04515}}].

\bibitem{Ben-David:2015b}
A.~{Ben-David}, S.~{von Hausegger} and A.D.~{Jackson}, \emph{{Skewness and kurtosis as indicators of non-Gaussianity in galactic foreground maps}}, \href{https://doi.org/10.1088/1475-7516/2015/11/019}{\emph{JCAP} {\bfseries 2015} (2015) 019} [\href{https://arxiv.org/abs/1509.03100}{{\ttfamily 1509.03100}}].

\bibitem{2019MNRAS.487.5814V}
S.~{von Hausegger}, A.~{Gammelgaard Ravnebjerg} and H.~{Liu}, \emph{{Statistical properties of polarized CMB foreground maps}}, \href{https://doi.org/10.1093/mnras/stz1582}{\emph{\mnras} {\bfseries 487} (2019) 5814} [\href{https://arxiv.org/abs/1811.02470}{{\ttfamily 1811.02470}}].

\bibitem{Rana:2018oft}
S.~{Rana}, T.~{Ghosh}, J.S.~{Bagla} and P.~{Chingangbam}, \emph{{Non-Gaussianity of diffuse Galactic synchrotron emission at 408 MHz}}, \href{https://doi.org/10.1093/mnras/sty2348}{\emph{Mon. Notices Royal Astron. Soc.} {\bfseries 481} (2018) 970} [\href{https://arxiv.org/abs/1806.01565}{{\ttfamily 1806.01565}}].

\bibitem{Waelkens:2008gp}
A.~{Waelkens}, T.~{Jaffe}, M.~{Reinecke}, F.S.~{Kitaura} and T.A.~{En{\ss}lin}, \emph{{Simulating polarized Galactic synchrotron emission at all frequencies. The Hammurabi code}}, \href{https://doi.org/10.1051/0004-6361:200810564}{\emph{Astron. Astrophys.} {\bfseries 495} (2009) 697} [\href{https://arxiv.org/abs/0807.2262}{{\ttfamily 0807.2262}}].

\bibitem{Thorne2017}
B.~Thorne, J.~Dunkley, D.~Alonso and S.~Næss, \emph{{The Python Sky Model: software for simulating the Galactic microwave sky}}, \href{https://doi.org/10.1093/mnras/stx949}{\emph{Monthly Notices of the Royal Astronomical Society} {\bfseries 469} (2017) 2821} [\href{https://arxiv.org/abs/https://academic.oup.com/mnras/article-pdf/469/3/2821/17638902/stx949.pdf}{{\ttfamily https://academic.oup.com/mnras/article-pdf/469/3/2821/17638902/stx949.pdf}}].

\bibitem{LaPorta2008}
L.~{La Porta}, C.~{Burigana}, W.~{Reich} and P.~{Reich}, \emph{{The impact of Galactic synchrotron emission on CMB anisotropy measurements. I. Angular power spectrum analysis of total intensity all-sky surveys}}, \href{https://doi.org/10.1051/0004-6361:20078435}{\emph{\aap} {\bfseries 479} (2008) 641} [\href{https://arxiv.org/abs/0801.0547}{{\ttfamily 0801.0547}}].

\bibitem{Planck2018IV}
{Planck Collaboration}, Y.~{Akrami}, M.~{Ashdown}, J.~{Aumont}, C.~{Baccigalupi}, M.~{Ballardini} et~al., \emph{{Planck 2018 results. IV. Diffuse component separation}}, \href{https://doi.org/10.1051/0004-6361/201833881}{\emph{\aap} {\bfseries 641} (2020) A4} [\href{https://arxiv.org/abs/1807.06208}{{\ttfamily 1807.06208}}].

\bibitem{Martire2022}
F.A.~{Martire}, R.B.~{Barreiro} and E.~{Mart{\'\i}nez-Gonz{\'a}lez}, \emph{{Characterization of the polarized synchrotron emission from Planck and WMAP data}}, \href{https://doi.org/10.1088/1475-7516/2022/04/003}{\emph{\jcap} {\bfseries 2022} (2022) 003} [\href{https://arxiv.org/abs/2110.12803}{{\ttfamily 2110.12803}}].

\bibitem{Chingangbam:2013}
P.~{Chingangbam} and C.~{Park}, \emph{{Residual foreground contamination in the WMAP data and bias in non-Gaussianity estimation}}, \href{https://doi.org/10.1088/1475-7516/2013/02/031}{\emph{JCAP} {\bfseries 2013} (2013) 031} [\href{https://arxiv.org/abs/1210.2250}{{\ttfamily 1210.2250}}].

\bibitem{bennett_nine-year_2013}
C.L.~Bennett, D.~Larson, J.L.~Weiland, N.~Jarosik, G.~Hinshaw, N.~Odegard et~al., \emph{{NINE}-{YEAR} \textit{{WILKINSON} {MICROWAVE} {ANISOTROPY} {PROBE}} ( \textit{{WMAP}} ) {OBSERVATIONS}: {FINAL} {MAPS} {AND} {RESULTS}}, \href{https://doi.org/10.1088/0067-0049/208/2/20}{\emph{The Astrophysical Journal Supplement Series} {\bfseries 208} (2013) 20}.

\bibitem{Planck2018I}
{Planck Collaboration}, N.~{Aghanim}, Y.~{Akrami}, F.~{Arroja}, M.~{Ashdown}, J.~{Aumont} et~al., \emph{{Planck 2018 results. I. Overview and the cosmological legacy of Planck}}, \href{https://doi.org/10.1051/0004-6361/201833880}{\emph{\aap} {\bfseries 641} (2020) A1} [\href{https://arxiv.org/abs/1807.06205}{{\ttfamily 1807.06205}}].

\bibitem{BP:Main}
{BeyondPlanck Collaboration}, K.J.~{Andersen}, R.~{Aurlien}, R.~{Banerji}, M.~{Bersanelli}, S.~{Bertocco} et~al., \emph{{BeyondPlanck I. Global Bayesian analysis of the Planck Low Frequency Instrument data}}, {\emph{arXiv e-prints} (2020) arXiv:2011.05609} [\href{https://arxiv.org/abs/2011.05609}{{\ttfamily 2011.05609}}].

\bibitem{1986A&AS...63..205R}
P.~{Reich} and W.~{Reich}, \emph{{A radio continuum survey of the northern sky at 1420 MHz. II}}, {\emph{Astron. astrophys., Suppl. ser. (Print)} {\bfseries 63} (1986) 205}.

\bibitem{testori_radio_2001}
J.C.~Testori, P.~Reich, J.A.~Bava, F.R.~Colomb, E.E.~Hurrel, J.J.~Larrarte et~al., \emph{A radio continuum survey of the southern sky at 1420 {MHz}: {Observations} and data reduction}, \href{https://doi.org/10.1051/0004-6361:20010088}{\emph{Astronomy \& Astrophysics} {\bfseries 368} (2001) 1123}.

\bibitem{Martire:2023ytg}
F.A.~Martire, A.J.~Banday, E.~Mart\'\i{}nez-Gonz\'alez and R.B.~Barreiro, \emph{{Morphological analysis of the polarized synchrotron emission with WMAP and Planck}}, \href{https://doi.org/10.1088/1475-7516/2023/04/049}{\emph{JCAP} {\bfseries 04} (2023) 049} [\href{https://arxiv.org/abs/2301.08041}{{\ttfamily 2301.08041}}].

\bibitem{WMAP:1yearFg}
C.L.~{Bennett}, R.S.~{Hill}, G.~{Hinshaw}, M.R.~{Nolta}, N.~{Odegard}, L.~{Page} et~al., \emph{{First-Year Wilkinson Microwave Anisotropy Probe (WMAP) Observations: Foreground Emission}}, \href{https://doi.org/10.1086/377252}{\emph{\apjs} {\bfseries 148} (2003) 97} [\href{https://arxiv.org/abs/astro-ph/0302208}{{\ttfamily astro-ph/0302208}}].

\bibitem{Orlando2013}
E.~{Orlando} and A.~{Strong}, \emph{{Galactic synchrotron emission with cosmic ray propagation models}}, \href{https://doi.org/10.1093/mnras/stt1718}{\emph{\mnras} {\bfseries 436} (2013) 2127} [\href{https://arxiv.org/abs/1309.2947}{{\ttfamily 1309.2947}}].

\bibitem{miville2008}
M.-A.~Miville-Deschênes, N.~Ysard, A.~Lavabre, N.~Ponthieu, J.F.~Macías-Pérez, J.~Aumont et~al., \emph{Separation of anomalous and synchrotron emissions using {WMAP} polarization data}, \href{https://doi.org/10.1051/0004-6361:200809484}{\emph{Astronomy \& Astrophysics} {\bfseries 490} (2008) 1093}.

\bibitem{2020MNRAS.495..578J}
L.~{Jew} and R.D.P.~{Grumitt}, \emph{{The spectral index of polarized diffuse Galactic emission between 30 and 44 GHz}}, \href{https://doi.org/10.1093/mnras/staa1233}{\emph{\mnras} {\bfseries 495} (2020) 578} [\href{https://arxiv.org/abs/1907.11426}{{\ttfamily 1907.11426}}].

\bibitem{fuskeland_constraints_2021}
U.~Fuskeland, K.J.~Andersen, R.~Aurlien, R.~Banerji, M.~Brilenkov, H.K.~Eriksen et~al., \emph{Constraints on the spectral index of polarized synchrotron emission from {WMAP} and {Faraday}-corrected {S}-{PASS} data}, \href{https://doi.org/10.1051/0004-6361/202037629}{\emph{Astronomy \& Astrophysics} {\bfseries 646} (2021) A69}.

\bibitem{Vidal2015}
M.~{Vidal}, C.~{Dickinson}, R.D.~{Davies} and J.P.~{Leahy}, \emph{{Polarized radio filaments outside the Galactic plane}}, \href{https://doi.org/10.1093/mnras/stv1328}{\emph{\mnras} {\bfseries 452} (2015) 656} [\href{https://arxiv.org/abs/1410.4438}{{\ttfamily 1410.4438}}].

\bibitem{Baccigalupi2015}
C.~{Baccigalupi}, C.~{Burigana}, F.~{Perrotta}, G.~{De Zotti}, L.~{La Porta}, D.~{Maino} et~al., \emph{{Power spectrum of the polarized diffuse Galactic radio emission}}, \href{https://doi.org/10.1051/0004-6361:20010414}{\emph{\aap} {\bfseries 372} (2001) 8} [\href{https://arxiv.org/abs/astro-ph/0009135}{{\ttfamily astro-ph/0009135}}].

\bibitem{Burigana2006}
C.~Burigana, L.~La~Porta, W.~Reich, P.~Reich, J.~Gonzalez-Nuevo, M.~Massardi et~al., \emph{{Polarized synchrotron emission}}, \href{https://doi.org/10.22323/1.027.0016}{\emph{PoS} {\bfseries CMB2006} (2006) 016} [\href{https://arxiv.org/abs/astro-ph/0607469}{{\ttfamily astro-ph/0607469}}].

\bibitem{Planck2018XI}
{Planck Collaboration}, Y.~{Akrami}, M.~{Ashdown}, J.~{Aumont}, C.~{Baccigalupi}, M.~{Ballardini} et~al., \emph{{Planck 2018 results. XI. Polarized dust foregrounds}}, \href{https://doi.org/10.1051/0004-6361/201832618}{\emph{\aap} {\bfseries 641} (2020) A11} [\href{https://arxiv.org/abs/1801.04945}{{\ttfamily 1801.04945}}].

\bibitem{Krachmalnicoff2018}
N.~{Krachmalnicoff}, E.~{Carretti}, C.~{Baccigalupi}, G.~{Bernardi}, S.~{Brown}, B.M.~{Gaensler} et~al., \emph{{S-PASS view of polarized Galactic synchrotron at 2.3 GHz as a contaminant to CMB observations}}, \href{https://doi.org/10.1051/0004-6361/201832768}{\emph{\aap} {\bfseries 618} (2018) A166} [\href{https://arxiv.org/abs/1802.01145}{{\ttfamily 1802.01145}}].

\bibitem{Strong2011}
A.W.~{Strong}, E.~{Orlando} and T.R.~{Jaffe}, \emph{{The interstellar cosmic-ray electron spectrum from synchrotron radiation and direct measurements}}, \href{https://doi.org/10.1051/0004-6361/201116828}{\emph{\aap} {\bfseries 534} (2011) A54} [\href{https://arxiv.org/abs/1108.4822}{{\ttfamily 1108.4822}}].

\bibitem{Cho:2010kw}
J.~Cho and A.~Lazarian, \emph{{Galactic foregrounds: Spatial fluctuations and a procedure of removal}}, \href{https://doi.org/10.1088/0004-637X/720/2/1181}{\emph{Astrophys. J.} {\bfseries 720} (2010) 1181} [\href{https://arxiv.org/abs/1007.3740}{{\ttfamily 1007.3740}}].

\bibitem{Lazarian2012}
A.~{Lazarian} and D.~{Pogosyan}, \emph{{Statistical Description of Synchrotron Intensity Fluctuations: Studies of Astrophysical Magnetic Turbulence}}, \href{https://doi.org/10.1088/0004-637X/747/1/5}{\emph{Astrophys. J.} {\bfseries 747} (2012) 5} [\href{https://arxiv.org/abs/1105.4617}{{\ttfamily 1105.4617}}].

\bibitem{Mertsch:2013pua}
P.~Mertsch and S.~Sarkar, \emph{{Loops and spurs: The angular power spectrum of the Galactic synchrotron background}}, \href{https://doi.org/10.1088/1475-7516/2013/06/041}{\emph{JCAP} {\bfseries 06} (2013) 041} [\href{https://arxiv.org/abs/1304.1078}{{\ttfamily 1304.1078}}].

\bibitem{Kendel2018}
D.~{Kandel}, A.~{Lazarian} and D.~{Pogosyan}, \emph{{Statistical properties of Galactic CMB foregrounds: dust and synchrotron}}, \href{https://doi.org/10.1093/mnras/sty1115}{\emph{\mnras} {\bfseries 478} (2018) 530} [\href{https://arxiv.org/abs/1711.03161}{{\ttfamily 1711.03161}}].

\bibitem{Haslam:1982}
C.G.T.~{Haslam}, C.J.~{Salter}, H.~{Stoffel} and W.E.~{Wilson}, \emph{{A 408 MHz all-sky continuum survey. II. The atlas of contour maps.}}, {\emph{Astron. astrophys., Suppl. ser. (Print)} {\bfseries 47} (1982) 1}.

\bibitem{Remazeilles:2014mba}
M.~{Remazeilles}, C.~{Dickinson}, A.J.~{Banday}, M.A.~{Bigot-Sazy} and T.~{Ghosh}, \emph{{An improved source-subtracted and destriped 408-MHz all-sky map}}, \href{https://doi.org/10.1093/mnras/stv1274}{\emph{Mon. Notices Royal Astron. Soc.} {\bfseries 451} (2015) 4311} [\href{https://arxiv.org/abs/1411.3628}{{\ttfamily 1411.3628}}].

\bibitem{WMAP:9year}
{\scshape WMAP} collaboration, \emph{{Nine-Year Wilkinson Microwave Anisotropy Probe (WMAP) Observations: Cosmological Parameter Results}}, \href{https://doi.org/10.1088/0067-0049/208/2/19}{\emph{Astrophys. J. Suppl.} {\bfseries 208} (2013) 19} [\href{https://arxiv.org/abs/1212.5226}{{\ttfamily 1212.5226}}].

\bibitem{Eriksen:2008}
H.K.~{Eriksen}, J.B.~{Jewell}, C.~{Dickinson}, A.J.~{Banday}, K.M.~{G{\'o}rski} and C.R.~{Lawrence}, \emph{{Joint Bayesian Component Separation and CMB Power Spectrum Estimation}}, \href{https://doi.org/10.1086/525277}{\emph{\apj} {\bfseries 676} (2008) 10} [\href{https://arxiv.org/abs/0709.1058}{{\ttfamily 0709.1058}}].

\bibitem{NaMaster:2019}
D.~Alonso, J.~Sanchez, A.~Slosar and L.D.E.S.~Collaboration, \emph{{A unified pseudo-$C_{\ell}$ framework}}, \href{https://doi.org/10.1093/mnras/stz093}{\emph{Monthly Notices of the Royal Astronomical Society} {\bfseries 484} (2019) 4127} [\href{https://arxiv.org/abs/https://academic.oup.com/mnras/article-pdf/484/3/4127/27747342/stz093.pdf}{{\ttfamily https://academic.oup.com/mnras/article-pdf/484/3/4127/27747342/stz093.pdf}}].

\bibitem{Drain2011_book}
B.T.~{Draine}, \emph{{Physics of the Interstellar and Intergalactic Medium}} (2011).

\bibitem{Planck_2015_X}
{Planck Collaboration}, R.~{Adam}, P.A.R.~{Ade}, N.~{Aghanim}, M.I.R.~{Alves}, M.~{Arnaud} et~al., \emph{{Planck 2015 results. X. Diffuse component separation: Foreground maps}}, \href{https://doi.org/10.1051/0004-6361/201525967}{\emph{\aap} {\bfseries 594} (2016) A10} [\href{https://arxiv.org/abs/1502.01588}{{\ttfamily 1502.01588}}].

\bibitem{Chingangbam:2017sap}
P.~{Chingangbam}, V.~{Ganesan}, K.P.~{Yogendran} and C.~{Park}, \emph{{On Minkowski Functionals of CMB polarization}}, \href{https://doi.org/10.1016/j.physletb.2017.05.030}{\emph{Physics Letters B} {\bfseries 771} (2017) 67} [\href{https://arxiv.org/abs/1705.04454}{{\ttfamily 1705.04454}}].

\bibitem{Chingangbam:2021}
P.~{Chingangbam}, P.~{Goyal}, K.P.~{Yogendran} and S.~{Appleby}, \emph{{Geometrical meaning of statistical isotropy of smooth random fields in two dimensions}}, \href{https://doi.org/10.1103/PhysRevD.104.123516}{\emph{\prd} {\bfseries 104} (2021) 123516} [\href{https://arxiv.org/abs/2109.05726}{{\ttfamily 2109.05726}}].

\bibitem{mcmullen1997isometry}
P.~McMullen, \emph{Isometry covariant valuations on convex bodies}, {\emph{Supplemento Ai Rendiconti Circ Mat Palermo} {\bfseries 50} (1997) 259}.

\bibitem{hug2008space}
D.~Hug, R.~Schneider and R.~Schuster, \emph{The space of isometry covariant tensor valuations}, {\emph{St. Petersburg Mathematical Journal} {\bfseries 19} (2008) 137}.

\bibitem{Schroder2D:2009}
G.~Schröder-Turk, S.~Kapfer, B.~Breidenbach, C.~Beisbart and K.~Mecke, \emph{Tensorial minkowski functionals and anisotropy measures for planar patterns}, \href{https://doi.org/https://doi.org/10.1111/j.1365-2818.2009.03331.x}{\emph{Journal of Microscopy} {\bfseries 238} (2010) 57}.

\bibitem{Tomita:1986}
H.~{Tomita}, \emph{{Curvature Invariants of Random Interface Generated by Gaussian Fields}}, \href{https://doi.org/10.1143/PTP.76.952}{\emph{Progress of Theoretical Physics} {\bfseries 76} (1986) 952}.

\bibitem{Matsubara:2003}
T.~{Matsubara}, \emph{{Statistics of Smoothed Cosmic Fields in Perturbation Theory. I. Formulation and Useful Formulae in Second-Order Perturbation Theory}}, \href{https://doi.org/10.1086/345521}{\emph{Astrophys. J.} {\bfseries 584} (2003) 1}.

\bibitem{Matsubara:2011}
T.~{Matsubara}, \emph{{Analytic Minkowski functionals of the cosmic microwave background: Second-order non-Gaussianity with bispectrum and trispectrum}}, \href{https://doi.org/10.1103/PhysRevD.81.083505}{\emph{Phys. Rev. D} {\bfseries 81} (2010) 083505} [\href{https://arxiv.org/abs/1001.2321}{{\ttfamily 1001.2321}}].

\bibitem{Matsubara:2020}
T.~{Matsubara} and S.~{Kuriki}, \emph{{Weakly non-Gaussian formula for the Minkowski functionals in general dimensions}}, {\emph{arXiv e-prints} (2020) arXiv:2011.04954} [\href{https://arxiv.org/abs/2011.04954}{{\ttfamily 2011.04954}}].

\bibitem{Gay:2012}
C.~{Gay}, C.~{Pichon} and D.~{Pogosyan}, \emph{{Non-Gaussian statistics of critical sets in 2D and 3D: Peaks, voids, saddles, genus, and skeleton}}, \href{https://doi.org/10.1103/PhysRevD.85.023011}{\emph{\prd} {\bfseries 85} (2012) 023011} [\href{https://arxiv.org/abs/1110.0261}{{\ttfamily 1110.0261}}].

\bibitem{Schmalzing:1998}
J.~{Schmalzing} and K.M.~{Gorski}, \emph{{Minkowski functionals used in the morphological analysis of cosmic microwave background anisotropy maps}}, \href{https://doi.org/10.1046/j.1365-8711.1998.01467.x}{\emph{Mon. Notices Royal Astron. Soc.} {\bfseries 297} (1998) 355} [\href{https://arxiv.org/abs/astro-ph/9710185}{{\ttfamily astro-ph/9710185}}].

\bibitem{Davies:2005za}
R.D.~Davies, C.~Dickinson, A.J.~Banday, T.R.~Jaffe, K.M.~Gorski and R.J.~Davis, \emph{{A determination of the spectra of galactic components observed by wmap}}, \href{https://doi.org/10.1111/j.1365-2966.2006.10572.x}{\emph{Mon. Not. Roy. Astron. Soc.} {\bfseries 370} (2006) 1125} [\href{https://arxiv.org/abs/astro-ph/0511384}{{\ttfamily astro-ph/0511384}}].

\bibitem{Ghosh:2012}
T.~{Ghosh}, A.J.~{Banday}, T.~{Jaffe}, C.~{Dickinson}, R.~{Davies}, R.~{Davis} et~al., \emph{{Foreground analysis using cross-correlations of external templates on the 7-year Wilkinson Microwave Anisotropy Probe data}}, \href{https://doi.org/10.1111/j.1365-2966.2012.20875.x}{\emph{\mnras} {\bfseries 422} (2012) 3617} [\href{https://arxiv.org/abs/1112.0509}{{\ttfamily 1112.0509}}].

\bibitem{Planck_2018III_HFI}
{Planck Collaboration}, N.~{Aghanim}, Y.~{Akrami}, M.~{Ashdown}, J.~{Aumont}, C.~{Baccigalupi} et~al., \emph{{Planck 2018 results. III. High Frequency Instrument data processing and frequency maps}}, \href{https://doi.org/10.1051/0004-6361/201832909}{\emph{\aap} {\bfseries 641} (2020) A3} [\href{https://arxiv.org/abs/1807.06207}{{\ttfamily 1807.06207}}].

\bibitem{Planck2015XXV}
{Planck Collaboration}, P.A.R.~{Ade}, N.~{Aghanim}, M.I.R.~{Alves}, M.~{Arnaud}, M.~{Ashdown} et~al., \emph{{Planck 2015 results. XXV. Diffuse low-frequency Galactic foregrounds}}, \href{https://doi.org/10.1051/0004-6361/201526803}{\emph{\aap} {\bfseries 594} (2016) A25} [\href{https://arxiv.org/abs/1506.06660}{{\ttfamily 1506.06660}}].

\bibitem{BP:Int_Fg}
K.J.~{Andersen}, R.~{Aurlien}, R.~{Banerji}, M.~{Bersanelli}, S.~{Bertocco}, M.~{Brilenkov} et~al., \emph{{BeyondPlanck XIV. Intensity foreground sampling, degeneracies and priors}}, {\emph{arXiv e-prints} (2022) arXiv:2201.08188} [\href{https://arxiv.org/abs/2201.08188}{{\ttfamily 2201.08188}}].

\bibitem{Watts:2023vdc}
D.J.~Watts et~al., \emph{{Cosmoglobe DR1 results. I. Improved Wilkinson Microwave Anisotropy Probe maps through Bayesian end-to-end analysis}},  \href{https://arxiv.org/abs/2303.08095}{{\ttfamily 2303.08095}}.

\bibitem{Carretti2019}
E.~Carretti et~al., \emph{{S-band Polarization All Sky Survey (S-PASS): survey description and maps}}, \href{https://doi.org/10.1093/mnras/stz806}{\emph{Mon. Not. Roy. Astron. Soc.} {\bfseries 489} (2019) 2330} [\href{https://arxiv.org/abs/1903.09420}{{\ttfamily 1903.09420}}].

\bibitem{Stuart2022}
S.E.~{Harper}, C.~{Dickinson}, A.~{Barr}, R.~{Cepeda-Arroita}, R.D.P.~{Grumitt}, H.M.~{Heilgendorff} et~al., \emph{{The C-Band All-Sky Survey (C-BASS): template fitting of diffuse galactic microwave emission in the northern sky}}, \href{https://doi.org/10.1093/mnras/stac1210}{\emph{\mnras} {\bfseries 513} (2022) 5900} [\href{https://arxiv.org/abs/2202.10411}{{\ttfamily 2202.10411}}].

\bibitem{QUIJOTEIV:2023}
J.A.~{Rubi{\~n}o-Mart{\'\i}n}, F.~{Guidi}, R.T.~{G{\'e}nova-Santos}, S.E.~{Harper}, D.~{Herranz}, R.J.~{Hoyland} et~al., \emph{{QUIJOTE scientific results - IV. A northern sky survey in intensity and polarization at 10-20 GHz with the multifrequency instrument}}, \href{https://doi.org/10.1093/mnras/stac3439}{\emph{\mnras} {\bfseries 519} (2023) 3383} [\href{https://arxiv.org/abs/2301.05113}{{\ttfamily 2301.05113}}].

\bibitem{Gorski:2005}
K.M.~{G{\'o}rski}, E.~{Hivon}, A.J.~{Banday}, B.D.~{Wandelt}, F.K.~{Hansen}, M.~{Reinecke} et~al., \emph{{HEALPix: A Framework for High-Resolution Discretization and Fast Analysis of Data Distributed on the Sphere}}, \href{https://doi.org/10.1086/427976}{\emph{Astrophys. J.} {\bfseries 622} (2005) 759} [\href{https://arxiv.org/abs/astro-ph/0409513}{{\ttfamily astro-ph/0409513}}].

\bibitem{Zonca:healpy}
A.~Zonca, L.~Singer, D.~Lenz, M.~Reinecke, C.~Rosset, E.~Hivon et~al., \emph{{healpy: equal area pixelization and spherical harmonics transforms for data on the sphere in Python}}, \href{https://doi.org/10.21105/joss.01298}{\emph{Journal of Open Source Software} {\bfseries 4} (2019) 1298}.

\bibitem{Hunter:2007}
J.D.~Hunter, \emph{Matplotlib: A 2d graphics environment}, \href{https://doi.org/10.1109/MCSE.2007.55}{\emph{Computing in Science \& Engineering} {\bfseries 9} (2007) 90}.

\end{thebibliography}\endgroup
\end{document}